\documentclass[a4paper]{jpconf}
\usepackage[super,compress]{cite}
\usepackage{graphicx}
\newcommand{\mrm}{\mathrm}
\newcommand{\half}{\mbox{$\frac{1}{2}$}}
\newcommand{\ra}{\mbox{$\rightarrow$}}
\begin{document}
\markboth{Dezs\H{o} Horv\'ath}
{Twenty Years of Searching for the Higgs boson}

%
%

\title{Twenty Years of Searching for the Higgs Boson:\\ 
Exclusion at LEP, Discovery at LHC}

\author{Dezs\H{o} Horv\'ath}

\address{Wigner Research Centre for Physics,
Budapest, Hungary
  \\ and {Atomki, Debrecen, Hungary}
 \\ horvath.dezso@wigner.mta.hu}


\begin{abstract}
The 40 years old Standard Model, the theory of particle physics, seems to
describe all experimental data very well. All of its elementary particles were
identified and studied apart from the Higgs boson until 2012. For decades many
experiments were built and operated searching for it, and finally, the two
main experiments of the Large Hadron Collider at CERN, CMS and ATLAS, in 2012
observed a new particle with properties close to those predicted for the Higgs
boson. In this paper we outline the search story: the exclusion of the Higgs
boson at LEP, the Large Electron Positron collider, and its observation at
LHC.

\end{abstract}


\section{Introduction: the Standard Model}
The Standard Model, the general theory of particle physics was established
more than 40 years ago. It describes our world as consisting of two kinds of
elementary particles, fermions and bosons, differing by their spin, intrinsic
angular momentum: fermions have half-integer, bosons have integer spins
measured in units of $\hbar$, the reduced Planck constant. The elementary
fermions have three families, each consisting of one pair of quarks and one
pair of leptons (Table \ref{tab:smpart}). Every fermion has an antiparticle of
the opposite charge. The leptons can propagate freely, but the quarks are
confined in hadrons: they can only exist in bound states of three quarks,
baryons (like the proton and neutron) or those of a quark and an antiquark,
mesons (like the pion). Three antiquarks make antibaryons like the antiproton.

\begin{table}
\begin{center}
\begin{tabular}{||l||c|c|c||c|cc||}
\hline
& Family 1 & Family 2 & Family 3 & charge & $T_3$&\multicolumn{1}{c||}{}\\
\hline
 &&&&&&\multicolumn{1}{c||}{}\\
Leptons & 
$\left(\begin{array}{l} \nu_{\mathrm e} \\ {\mathrm e} \end{array} 
\right)_L$ &
$\left(\begin{array}{l} \nu_{\mu} \\ \mu \end{array} \right)_L$ &
$\left(\begin{array}{l} \nu_{\tau} \\ \tau \end{array} \right)_L$
& $\begin{array}{r} ~0 \\ -1 \end{array} $
& $\begin{array}{r} +\frac{1}{2} \\ -\frac{1}{2} \end{array} $&\multicolumn{1}{c||}{}\\
 &&&&&&\multicolumn{1}{c||}{}\\
\hline
 &&&&&&\multicolumn{1}{c||}{}\\
Quarks &
$\left(\begin{array}{l} {\mathrm u} \\ {\mathrm d^{\prime}} \end{array} 
\right)_L $ &
$\left(\begin{array}{l} {\mathrm c} \\ {\mathrm s^{\prime}} \end{array} 
\right)_L $ &
$\left(\begin{array}{l} {\mathrm t} \\ {\mathrm b^{\prime}} \end{array} 
\right)_L$ 
& 
$\begin{array}{r} +\frac{2}{3} \\ -\frac{1}{3} \end{array} $
& $\begin{array}{r} +\frac{1}{2} \\ -\frac{1}{2} \end{array} 
$&\multicolumn{1}{c||}{}\\
 &&&&&&\multicolumn{1}{c||}{}\\
\hline
\end{tabular}
\caption[Leptons and quarks]{\label{tab:smpart}Leptons and quarks, the three
  families of basic fermions. $T_3$ is the third component of the weak
  isospin, the rest of the notation is explained in the text step by step.}
\label{tab:smpart}
\end{center}
\end{table}

In the Standard Model the three basic particle interactions, the strong
interaction holding the quarks in the nucleons and the nucleons in the atomic
nucleus, the weak interaction, responsible for the decay of the heavier basic
fermions, and the well known electromagnetic interaction, are all derived from
local gauge symmetries. A gauge symmetry is a freedom to define the coordinate
system measuring the strength of an interaction, the best known example of
which is the freedom to choose the potential zero of an electric field. A
local symmetry is its modified form when the gauge is changing in space-time
according to a known function. The three basic interactions are mediated by
elementary bosons: the strong nuclear force by 8 gluons, the weak interactions
by the three heavy weak bosons and the electromagnetism by the photon. In
order to cancel uncomfortable terms from equations the theory also needs the
existence of an additional scalar boson, a particle with all its quantum
numbers like charges and spin zero.

Local gauge symmetries give correct answers to important questions except the
mass of elementary particles: one has to violate them in order to get non-zero
masses. This spontaneous symmetry breaking (SSB) mechanism was
introduced in several steps to particle physics and it is now an integral part
of the Standard Model. It is called, somewhat unjustified, also the Higgs
mechanism, although it is the product of several people, so it could also be
called {\em Brout--Englert--Higgs--Guralnik--Hagen--Kibble} mechanism
\cite{ref:Englert,ref:HiggsA,ref:HiggsB,ref:HiggsC,ref:Guralnik}. These days
we try to call it BEH mechanism by the initials of those who have first
published it. As the scalar boson was first introduced by Peter Higgs, it is
justly called Higgs boson.

The spontaneous symmetry breaking mechanism consists of adding to the
electroweak vacuum a potential which breaks its perfect symmetry. The form of
this potential is
\begin{equation}
V=\mu^2 \Phi^\dagger\Phi + \lambda (\Phi^\dagger\Phi)^2
\label{eq:beh-pot}
\end{equation}
where $\Phi$ is a complex doublet field of four components and $\lambda >0$ is
a real constant. If $\mu^2 >0$ then it is a scalar field with a non-zero mass
$\mu$ and a potential minimum at $\Phi=0$. However, if $\mu^2 <0$ then it has
a non-zero minimum
\begin{equation}
\Phi^\dagger\Phi =\half (\Phi_1^2 +\Phi_2^2  +\Phi_3^2  + \Phi_4^2)=
- \frac{\mu^2}{2\lambda}.
\end{equation}
$\Phi$ can be chosen so that it had just one real component and then expanded around its minimum, {\em vacuum expectation value}
$v=\sqrt{-\frac{\mu^2}{\lambda}}$ as
\begin{equation}
\Phi(x)= \sqrt{\half}\left(\begin{array}{c}0\\v+h(x)\end{array}\right)
\label{eq:BEH-doublet}
\end{equation}
The $\Phi$ field is imposed onto the vacuum of the $U(1)_Y\otimes SU(2)_L$
combined local gauge symmetries, breaking them. Here 
\begin{equation}
Y = 2 (Q-T^3) 
\label{eq:hypch}\end{equation}
the hypercharge of the fermion (where $Q$ is the electric charge and $T^3$ is
the third component of the weak isospin, see Table~\ref{tab:smpart}), and
$SU(2)_L$ acts on left-polarized fermion doublets. As the photon has zero
mass, from the neutral part of the electroweak Lagrangian we separate a good
$U(1)_Q$ using relation (\ref{eq:hypch}) between electric charge $Q$ and
hypercharge $Y$, and the rest gives a correct weak interaction with three
massive gauge bosons, W$^\pm$ and Z.  As a result of this manipulation, 3
components of the $\Phi$ field become the longitudinal polarizations (masses)
of the 3 $SU(2)$ gauge bosons and the fourth component makes the scalar Higgs
boson. The charged currents of the weak interaction, mediated by W$^\pm$ are
pure $SU(2)_L$ whereas Z is the result of the unification, thus they have
different masses and Z mediates a right-handed neutral current as well.

\begin{figure}
\begin{center}
\includegraphics[width=0.5\linewidth]{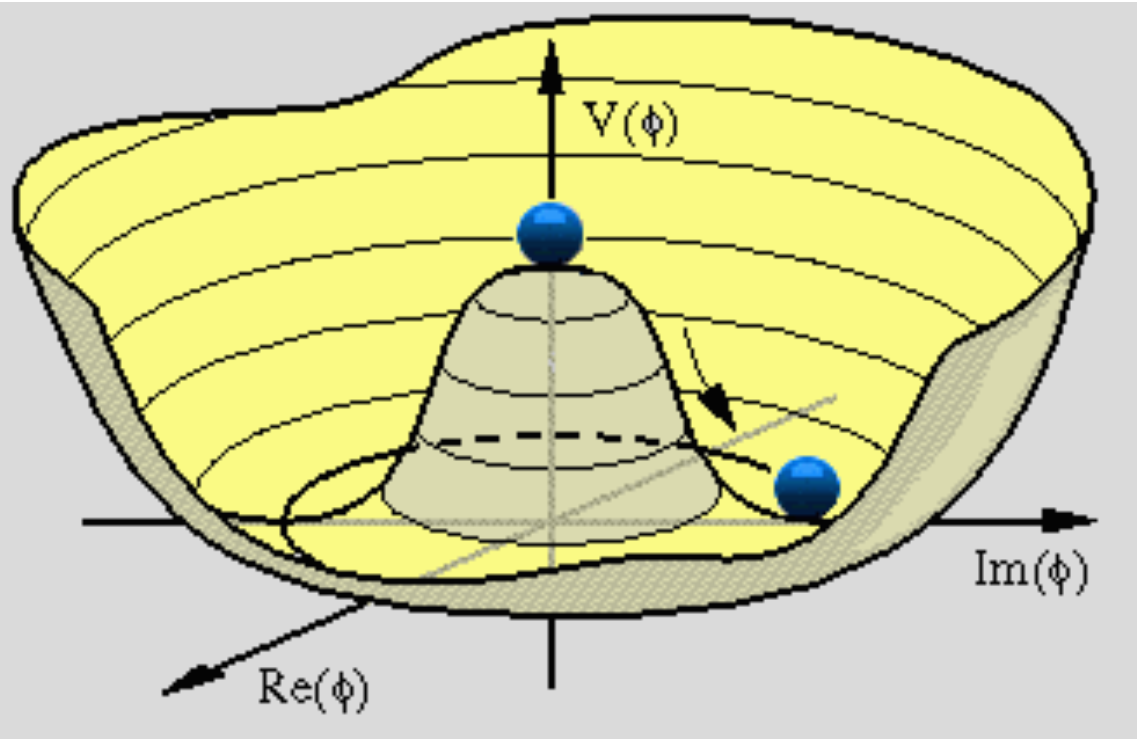}
\end{center}
\caption{\label{fig:ssb}Spontaneous symmetry breaking: the BEH
  potential. The axial symmetry of the potential is not violated by putting a
  ball on the top at $\Phi=0$, but it will be spontaneously broken when the ball
  rolls down in the valley. However, the coordinate system can always be
  chosen so that the ball were at point $Im(\Phi)=0$.}
\end{figure}

Spontaneous symmetry breaking is well illustrated by a Mexican hat
(Fig. \ref{fig:ssb}). Its axial symmetry is not violated by putting a ball on
its top, however, the ball will eventually go down and break the original
symmetry. SSB makes it possible to introduce masses in the theoretical
equations: masses for the heavy weak bosons, W$^+$, W$^-$ and Z$^0$ mediating
the weak interaction and also masses for the basic fermions, the quarks and
leptons. There is a substantial difference, however: whereas the masses of the
weak bosons appear as a result of SSB, and so they are predicted by the
Standard Model, the fermion mass terms have to be added ad hoc to the Standard
Model Lagrangian as a triple interaction among the left-polarized fermion
doublet, the (\ref{eq:BEH-doublet}) BEH doublet and the right-polarized fermion
singlet field with arbitrary coupling constants giving the fermion masses. The
BEH field introduces 2 free parameters in the Standard Model; as the vacuum
expectation value can be determined from the Fermi coupling constant 
($v\sim246$ GeV), the only adjustable parameters left in the electroweak
theory are the $U(1)$ and $SU(2)$ coupling constants and the masses of the
fermions and of the Higgs boson.

Note that the masses of our macroscopic world are mostly due to the energy
content of the proton and neutron and not due to the BEH mechanism.

\begin{figure}
\begin{center}
\includegraphics[width=0.7\linewidth]{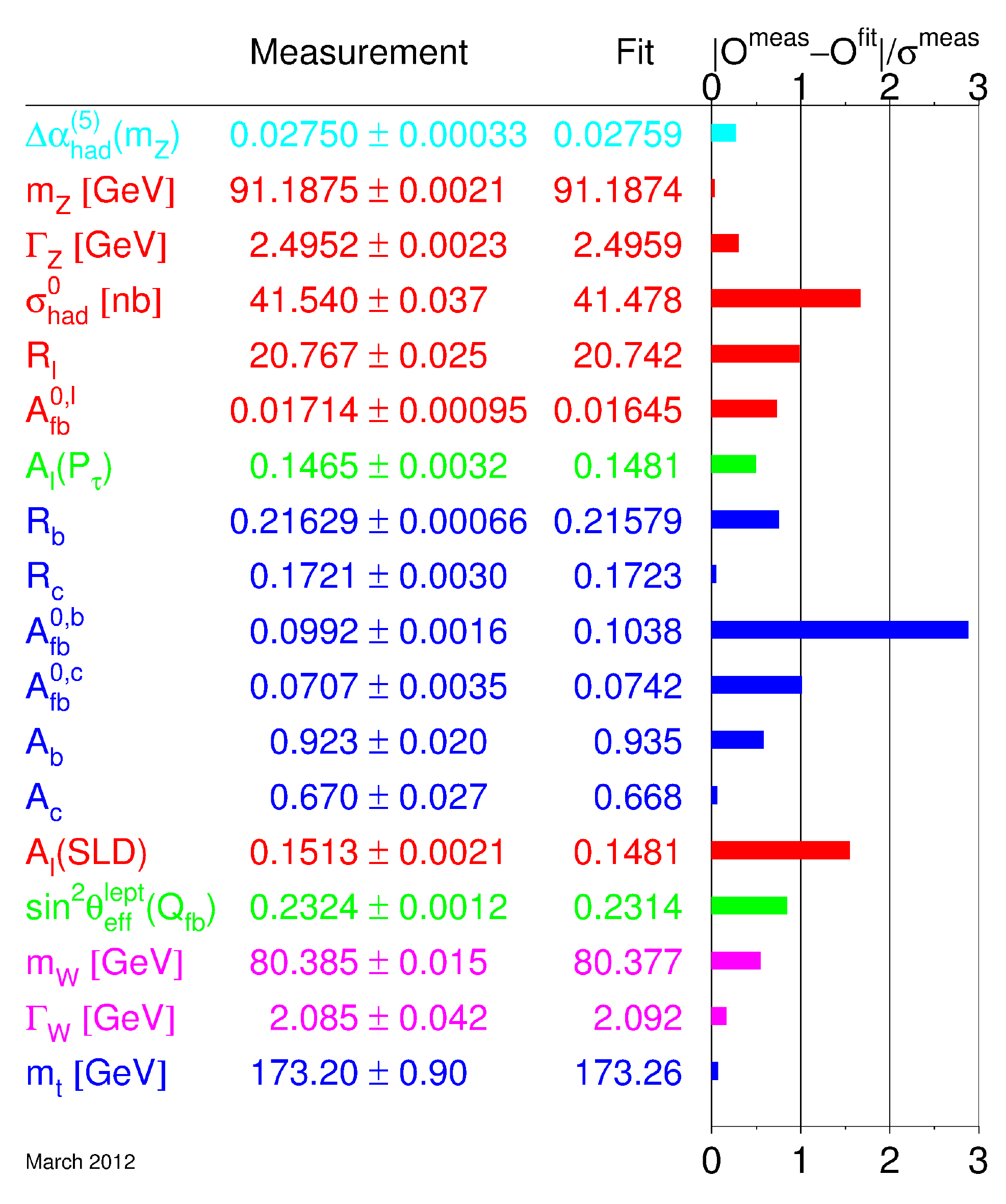}
\end{center}
\caption{\label{fig:show_pull} Various parameters of the Standard Model
  \cite{ref:ewwg}: average measured values (2nd column) with uncertainties
  (3rd column), the predictions or fits by the Standard Model (4th column) and
  a bar plot showing the differences between theory and experiment divided by
  the experimental uncertainties. The agreement is purely statistical as the
  difference is in only one case more than 2 uncertainties.}
\end{figure}

Since almost 40 years, more and more precise new data were acquired at the
particle accelerators and all seem to agree very well with the predictions of
the Standard Model. Hundreds of experiments are summarized in the few
parameters of Fig.~\ref{fig:show_pull} according to the LEP Electroweak
Working Group \cite{ref:ewwg}. It shows the 2012 situation of the analysis of
electroweak data: all experimental data and theoretical estimates agree within
the statistical boundaries. The only parameter which deviates at more than 2
uncertainties is the forward-backward asymmetry of the decay of the Z boson to
two b quarks.

The Higgs boson of the Standard Model is the only scalar particle: all of its
quantum numbers are zero, its only property is mass. Fitting experimental data
predicts that the Higgs mass should be around 100 GeV (between 80 and 160 GeV
within 95\% confidence). All constituents of the Standard Model were
identified and studied experimentally before the launch of the LHC, apart from
the Higgs boson, that is how it became the most wanted particle. As Peter
Higgs himself told \cite{ref:Higgs2002} ``It was in 1972 ... that my life as
a boson really began''.

\section{Analysis Concepts of Particle Physicists}

The statistical methods\footnote{This section follows the line of
ref.\ \cite{ref:hdstat}} used by particle physicists are as different from the
{\em official} mathematical statistics as mechanical engineering from the
Lagrangian or Hamiltonian formulation of theoretical mechanics. At the same
time statistics is extremely important for data analysis in particle physics:
every few years international workshops are organized by particle physicists
working at the Large Hadron Collider to exchange ideas on statistical methods,
the last one having been in 2011 \cite{ref:phys-stat-2011}.  In the Appendix
of that volume {\em Eilam Gross} defines the aim of his paper {\em LHC
  Statistics for Pedestrians}: {\em ''A pedestrian's guide $\ldots$ to help
  the confused physicist to understand the jargon and methods used by HEP
  phystatisticians.  $\ldots$ A phystatistician is a physicist who knows his
  way in statistics and knows how Kendall's advanced theory of statistics book
  looks like.}

Every high-energy collaboration has phystatistician experts and they all have
quite different ideas how to analyze data. In order to avoid confusion, the
large LHC collaborations have Statistics Committees which publish home pages
of recommendations how to do things. The Statistics Committees of both CMS and
ATLAS have several members who published text books on statistics for
physicists and ATLAS and CMS have a joint such committee as well.

As in high energy physics the primary experimental data are basically event
counts, the basic concepts are Poisson-like. The data follow the Poisson
distribution ($n_i$ events in bin $i$): ~ ${\cal P}(n_i|\mu_i) =
\frac{\mu_i^{n_i} e^{-\mu_i}}{n_i !}$ and the result is usually expressed in
terms of the Poisson likelihood: ${\cal L} = \Pi_i {\cal P}(n_i|\mu_i)$. 
The expected number of events is $\tilde\mu_i = \sum_{j} L \sigma_j
\epsilon_{ji}$, where $L$ is the integral luminosity collected, $\sigma_j$ is the
cross section of source $j$ and $\epsilon_{ji}$ is the efficiency (determined
by Monte Carlo simulation) of source $j$ in bin $i$.

Luminosity is the rate of collecting data for colliders, similar to the flux
of fixed-target experiments. It is defined as $L = f n \frac{N_1 N_2}{A}$
where $f$ is the circulation frequency of the colliding beams; $n$ is the
number of particle bunches in the ring; $N_1, N_2$ are the numbers of
particles in the two kinds of bunches; $A$ is the spatial overlap of the
colliding bunches. The total number of collisions is characterized by the
integrated luminosity: $\int_{t_1}^{t_2} L dt$ which is usually measured in
units of inverse cross-section, at LHC in $[\mrm{pb^{-1},\; fb^{-1}}]$. The
expected rate of a reaction with cross section $\sigma$ at $\epsilon$
detection efficiency is $R=\epsilon\sigma L$.

According to the general convention in accelerator experiments a given new
phenomenon is excluded if we can show it not appearing at a $\geq95 \%$
confidence level and observed if it exceeds $>5\sigma$ above background where
now, for a change, $\sigma$ is the experimental uncertainty according to the
best honest guess of the experimentalist.

That $\sigma$ uncertainty has a {\em statistical component} from the number of
observed events and {\em systematic} ones from various sources, like Monte
Carlo statistics and inputs, experimental calibration factors, detection
efficiencies, etc, with the common name {\em nuisance parameters}. To get a
rough estimation of the total error the systematic uncertainties could be
added quadratically to the statistical one: $\sigma= \sqrt{\sigma_\mrm{stat}^2
  + \sigma_\mrm{syst}^2}$, but in practice we derive the final uncertainty via
marginalizing (integrating out) \cite{ref:cous-high} the nuisance parameters
$\Theta$ in likelihood ${\cal L}$ using the related probability distributions
${\cal W}$: ${\cal L}(P;x) = {\cal W}(x|P) = \int {\cal W}(x|P,\Theta) {\cal
  W}(\Theta|P) d\Theta$.

Another important feature of high-energy data analysis is the {\em blind
  analysis} \cite{ref:blind}: ``A blind analysis is a measurement which is
performed without looking at the answer. Blind analyses are the optimal way to
reduce or eliminate experimenter's bias, the unintended biasing of a result in
a particular direction.'' It came from medical research and the idea is to
optimize, prove and publish your analysis technique using simulations and
earlier data only before touching new data in the critical region. For
instance, in Spring and early Summer, the 2012 CMS data were blinded in the
invariant mass region $110< M_H < 140$ GeV (where $M_H$ is the simulated Higgs
mass) because of the $3\sigma$ excess observed in 2011. The same procedure was
used again in Autumn 2012. The methods had to be fixed and approved by the
collaboration before simultaneous {\em unblinding} for all analysis channels.

\section{Search for the Higgs boson}

What we usually try to observe is a resonance. For a particle with lifetime
$\tau=\Gamma^{-1}$ and decay rate $\Gamma$ the event rate against the
invariant mass of the decay products is $|\chi(E)|^2
=\frac{1}{(E-M)^2+\Gamma^2/4}$, i.e. a Lorentz curve (Breit-Wigner
resonance). It shows a peak at the $M$ invariant mass of the decaying system
with a full width at half maximum $\Gamma$. The discovery of a new particle
can be claimed if we see a resonance at the same invariant mass of the
particle in all expected decay channels, by all related experiments. Many
hopeful new observations were disproved as statistical fluctuations by other
experiments, and a few stayed unproven as not confirmed. 

The search involves several consecutive steps.
\begin{itemize}
\item Compose a complete {\em Standard Model background} using Monte Carlo
  simulation taking into account all types of possible events normalized to
  their cross-sections.
\item Compose {\em Higgs signals}, simulations of all possible production and
  decay processes with all possible Higgs-boson masses.
\item Put all these through the {\em detector simulation} to get events
  analogous to the expected measured ones.
\item {\em Optimize the event selection} via reducing the  $B$  background and 
enhancing the $S$ signal via maximizing e.g.\ 
$N_S/\sqrt{N_B}$ or \cite{ref:BitKras} $2\cdot(\sqrt{N_S+N_B}-\sqrt{N_B})$.
\item {\em Check the background}, i.e.\ the description of data by the
  simulation for the given luminosity: the simulation should reproduce the
  observed background distributions in all details. For instance, you can
  check the background of the decay of a neutral particle to charged leptons
  by selecting lepton pairs of identical charges.
\item {\em Check the signal:} does it agree with the expectation by the
  theoretical model?
\end{itemize}

Once you are happy with the simulations and the event selection, you must
chose a test statistic. That could be any kind of probability variable
characteristic of the given phenomenon: probabilities for having background
only, for having signal or combinations. One of the favorite is the $Q$
likelihood ratio of signal + background over background: $Q = {\cal
  L}_{s+b}/{\cal L}_{b}$. As you see, although our basic approach is
definitely frequentist there is a certain Bayesian influence as well.  What
most frequently plotted is

$-2 \ln
Q(m_H) = 2 \sum_{k=1}^{N_\mrm{ch}} \left[s_k(m_H) - \sum_{j=1}^{n_k}
  \ln\left(1+\frac{s_k(m_H)S_k(x_{jk};m_H)}{b_kB_k(x_{jk})}\right) \right]$

Here the variables are the following:
\begin{itemize}
\item $n_k$: events observed in channel $k$, ~~ $k=1\ldots N_\mrm{ch}$.
\item $s_k(m_H)$ and $b_k$: signal and background events in channel $k$ for
  Higgs mass $m_H$.
\item $S_k(x_{jk};m_H)$ and $B_k(x_{jk})$: probability distributions for
  events for Higgs mass $m_H$ at test point $x_{jk}$.
\item $x_{jk}$: position of event $j$ of channel $k$ on the plane of its
  reconstructed Higgs mass and cumulative testing variable constructed of
  various special features of the event like b-tagging, signal likelihood,
  neural network output, etc.
\end{itemize}

Several other testing variables can be constructed on the same basis, the most
frequently used ones are probabilities of NOT having the expected signal on
the basis of the expected background and the collected data:
\begin{itemize}
\item $CL_b$, the signal confidence level assuming background only, i.e. the
  complete absence of the signal, or
\item The so-called {\em $p-$value}: the probability of obtaining a test
  statistic at least as extreme as the one that was actually observed,
  assuming that the null hypothesis is true. Translated to our language that
  means the probability that random fluctuation of the measured background
  could give the observed excess (if any).
\end{itemize}

\section{Exclusion at LEP}

\begin{figure}[h] 
\centering
\includegraphics[width=0.7\linewidth]{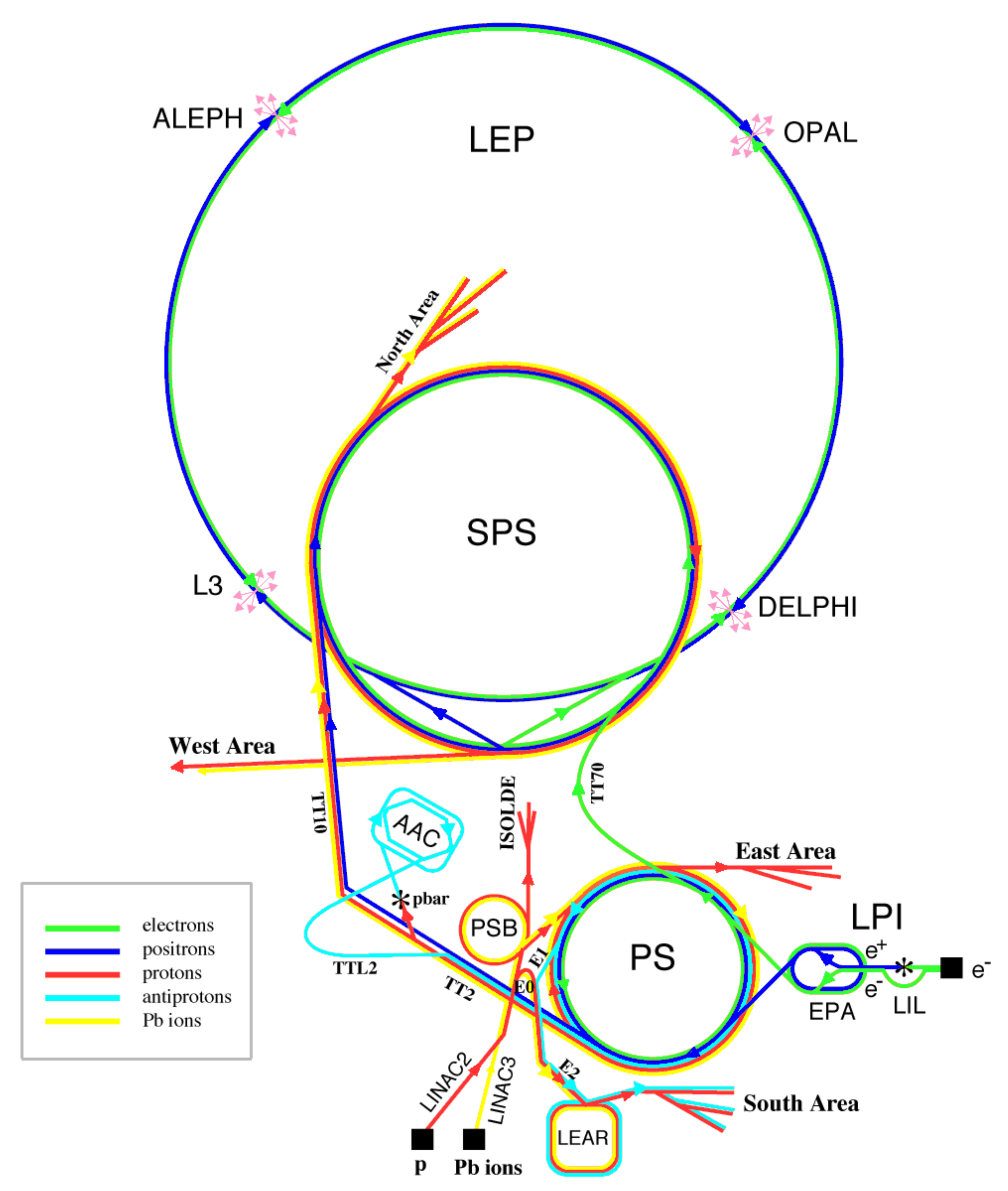}
\caption[The accelerator complex of CERN in the LEP era]{The accelerator
  complex of CERN in the LEP era}
\label{fig:LEP}
\end{figure}

Although the experiments of the Large Electron Positron (LEP) collider
(Fig.~\ref{fig:LEP}) saw no earthshaking discoveries, no real deviation from
the Standard Model, LEP provided an incredible amount of very precise
measurements, some of which are presented in Fig.~\ref{fig:show_pull}. In its
last two years of working, LEP was mostly devoted to the search for the Higgs
boson, collecting more luminosity at higher energies than in the previous 10
years together.

\begin{figure}[h]
\begin{center}
\includegraphics[width=0.6\linewidth]{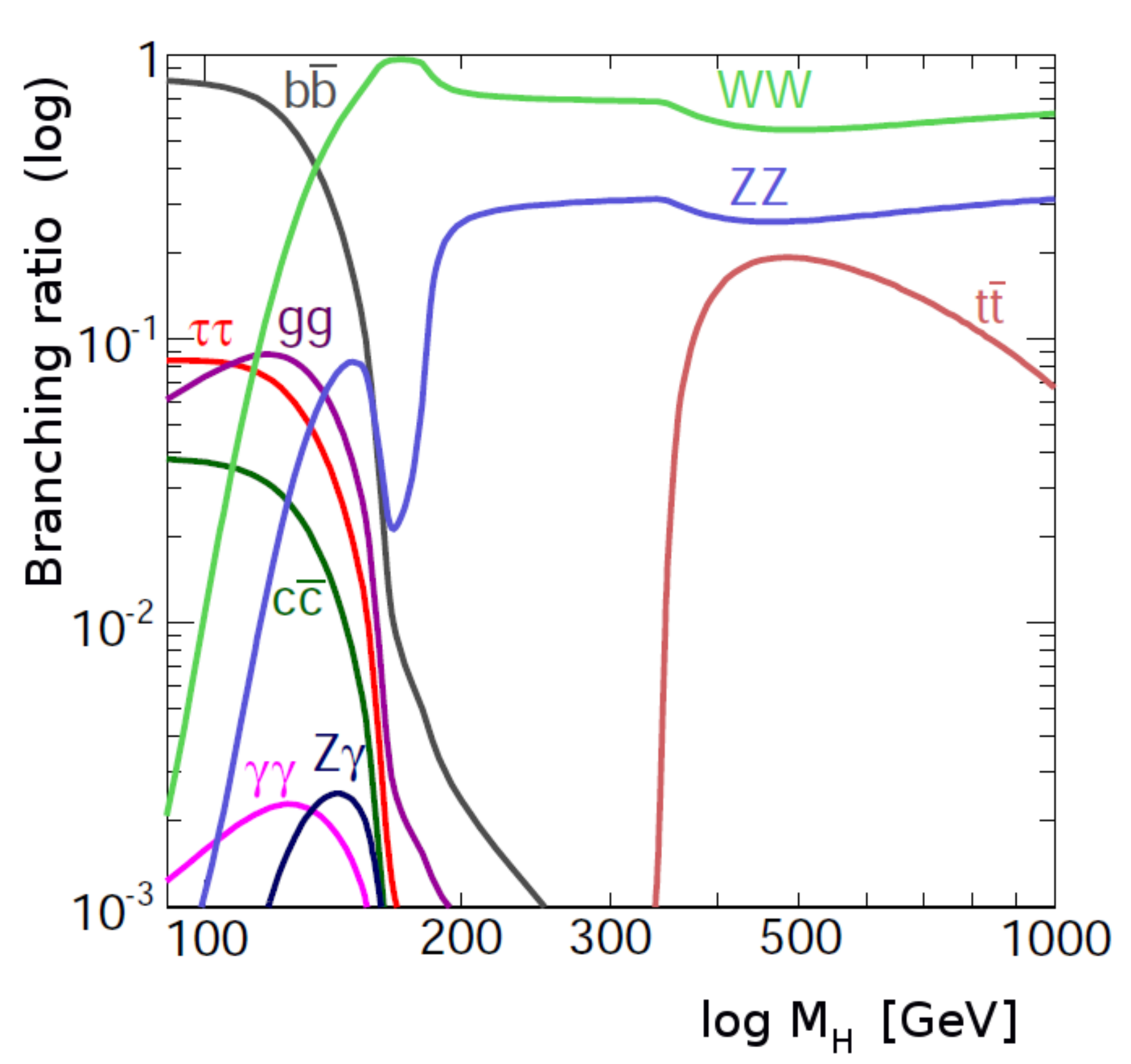}
\caption{\label{fig:SMH_br1}The various decay channels of the Higgs boson
  according to the Standard Model. Below 120~GeV the $\mathrm{H
    \ra\ b\overline{b}}$ decay channel dominates.}
\end{center}
\end{figure}

\begin{figure}[h]
\begin{center}
\includegraphics[width=0.9\linewidth]{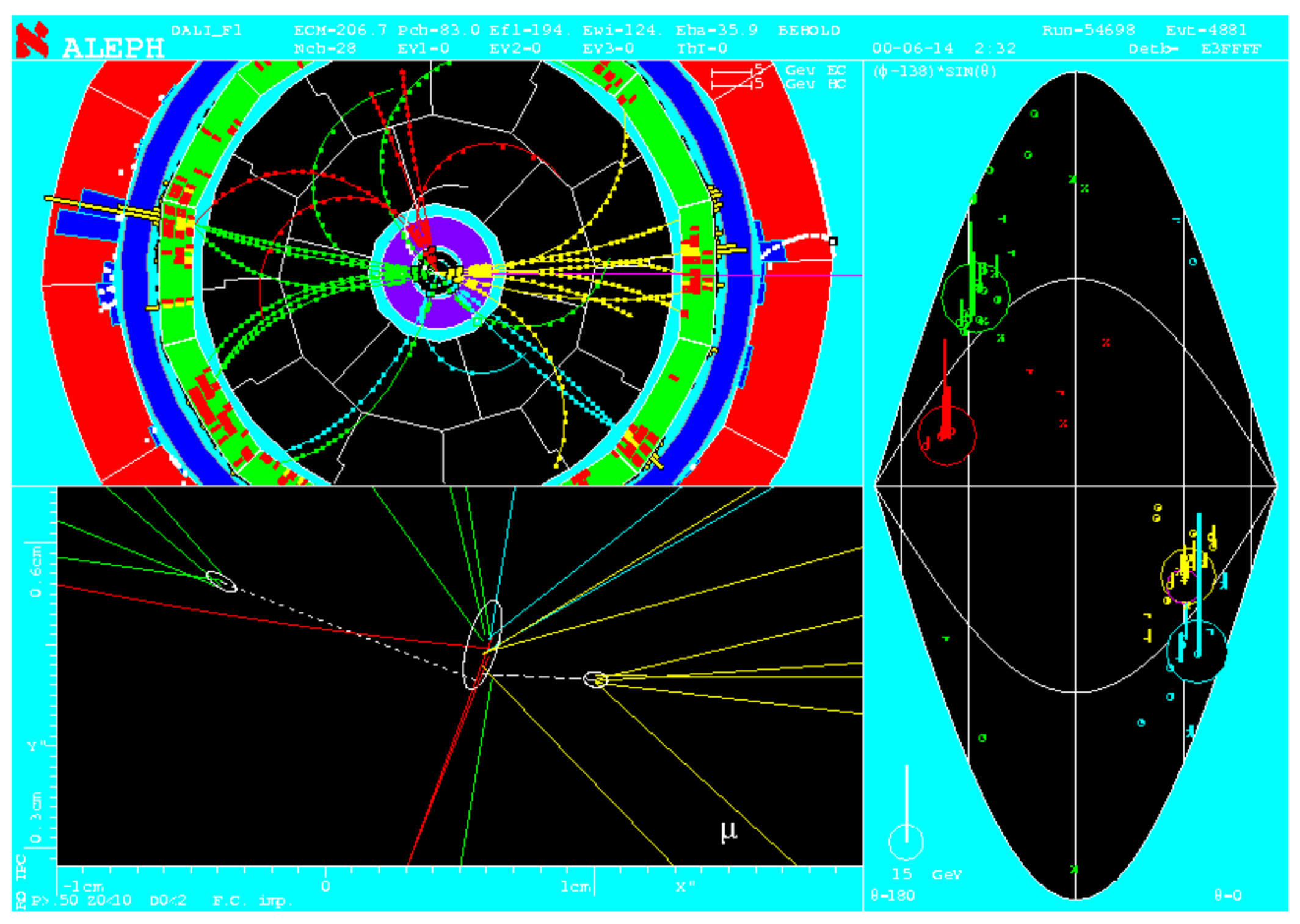}
\caption{\label{fig:aleph-event}Higgs-like event detected by the ALEPH
  experiment at LEP: an $\mrm{e^+e^-}$ collision produces 4 hadrons containing
  b quarks recognized by their secondary vertices due to longer lifetimes.}
\end{center}
\end{figure}

At LEP the dominant formation process is {\em Higgs-strahlung} $\mrm{e^-e^+\ra
  ZH}$ (the name comes from the funny {\em English} word
Bremsstrahlung\footnote{Other languages use the simple mirror translation of
  the German word: {\em braking radiation}}) and the dominant Higgs decay is
to 2 b-quarks. The various channels are different only due to the various
decay processes of the accompanying Z boson.

LEP had 4 large experiments (Fig.~\ref{fig:LEP}) in the 4 interaction points
of the electron-positron collider, ALEPH, DELPHI, L3 and OPAL (the present
author was working in OPAL). The structure of all contemporary high-energy
detectors are very similar, consisting of onion-like layers. A sensitive pixel
detector right around the beam pipe, a tracking system of multiwire chambers
or semiconductor detectors of minimal weight material following the tracks of
charged particles in the magnetic field of the detector, then an electromagnetic
calorimeter, something heavy absorbing all electrons and photons, outside of
that an even heavier hadron calorimeter, absorbing the pions, protons,
neutrons, etc., and finally, muon chambers, identifying the path of energetic
muons leaving the system. All detectors have huge magnets encompassing as much
as possible of the detector parts. Figure \ref{fig:aleph-event} shows a
Higgs-boson-like event detected by ALEPH in the 4-jet channel.

\begin{figure}[h]
\begin{center}
\includegraphics[width=\linewidth]{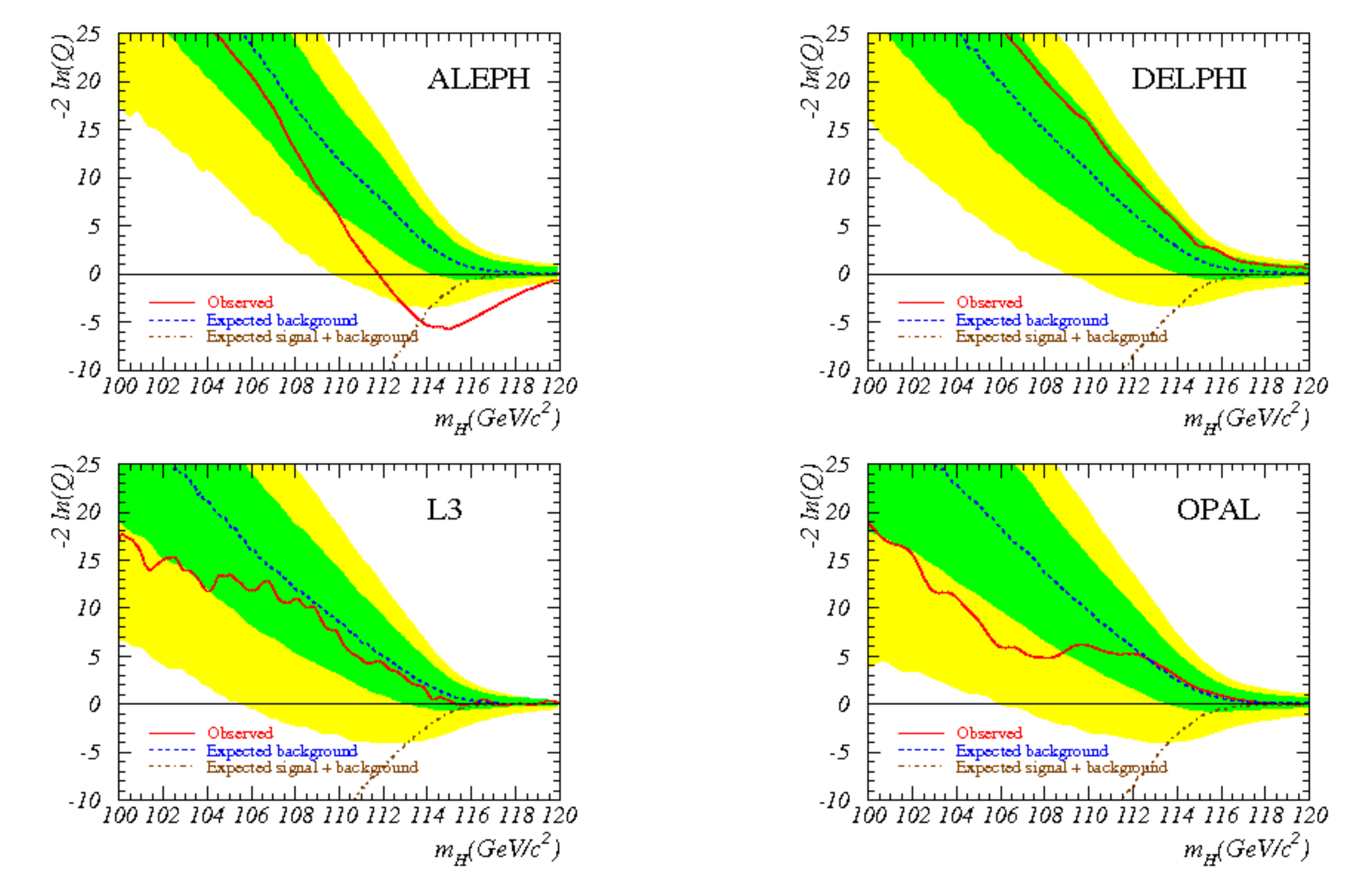}
\end{center}
\caption{\label{fig:adlo_2lnq}Exclusion of the Higgs boson at LEP. The test
  statistic, $-2\ln{Q}$ shows a significant signal for ALEPH and nothing for
  the other 3 LEP experiments at equivalent statistical and experimental
  circumstances. The observed signal of ALEPH by far exceeds the expectations
  of the Standard Model.
}
\end{figure}

Statistics played a rough joke at LEP: one of the experiments, ALEPH, saw in
one of the possible Higgs decay channels a very significant signal
corresponding to a Higgs boson of a mass of 115~GeV/$c^2$, while the rest of
LEP have not seen anything \cite{ref:lephiggs2003}
(Fig.~\ref{fig:adlo_2lnq}). ALEPH saw the excess in 4-jet events only, in
those events where the Higgs boson decays to a pair of b quarks and the
accompanying Z boson also decays to a quark pair
(Fig.~\ref{fig:aleph-event}). The b quark is identified by its long lifetime
leading to a secondary decay vertex in the event.  Another strange thing was
that the Higgs signal seen by ALEPH by far exceeded the expectations of the
Standard Model. Also, the observed Higgs mass was critical as it coincided
with the average kinematic limit of LEP: in 2000 the average collision energy
of LEP was about 206 GeV and the observed resonance was found at
115~GeV/$c^2$, the difference is very close to the mass of the Z boson,
91~GeV/$c^2$. 

\begin{figure}[h]
\begin{center}
\includegraphics[width=0.35\linewidth]{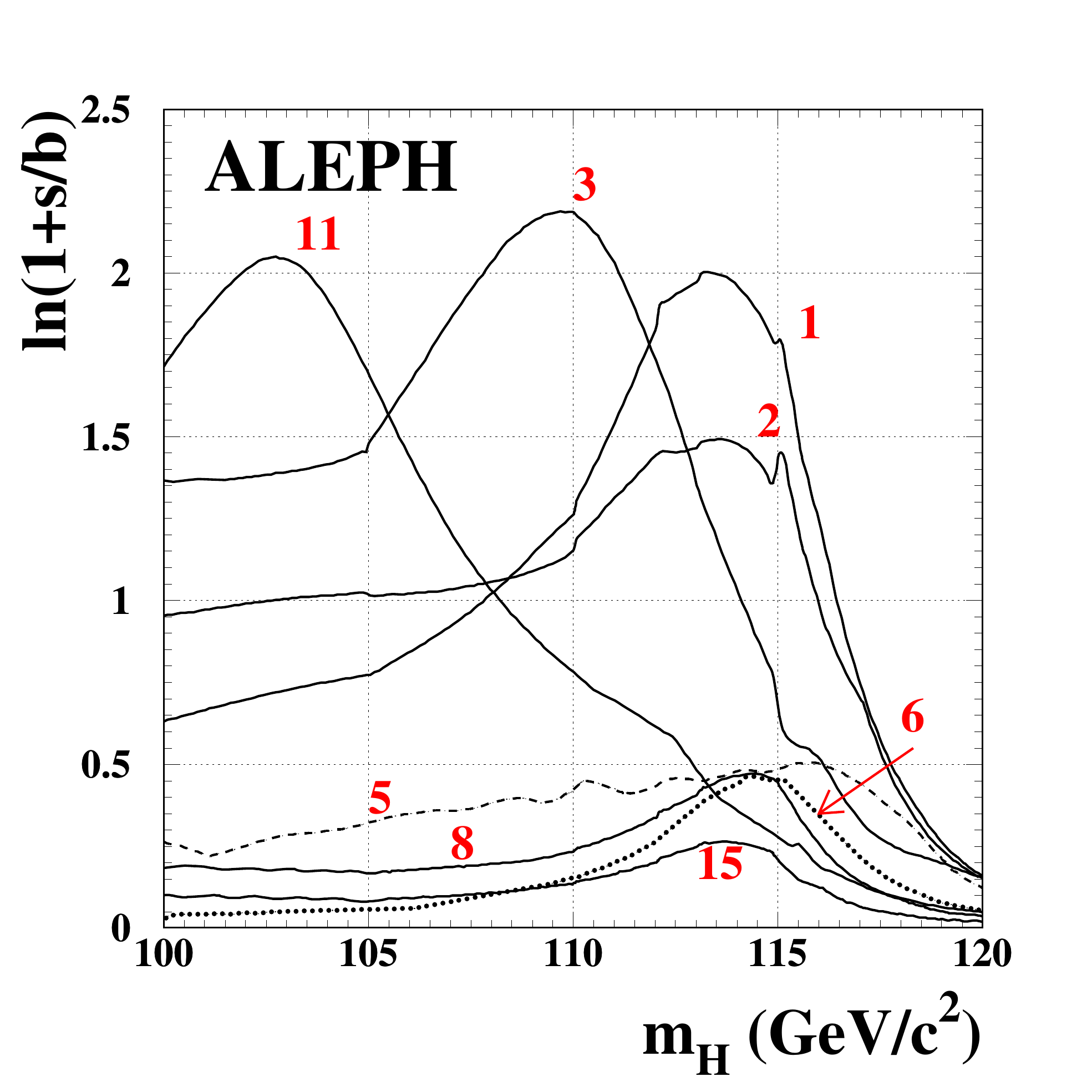} 
\includegraphics[width=0.35\linewidth]{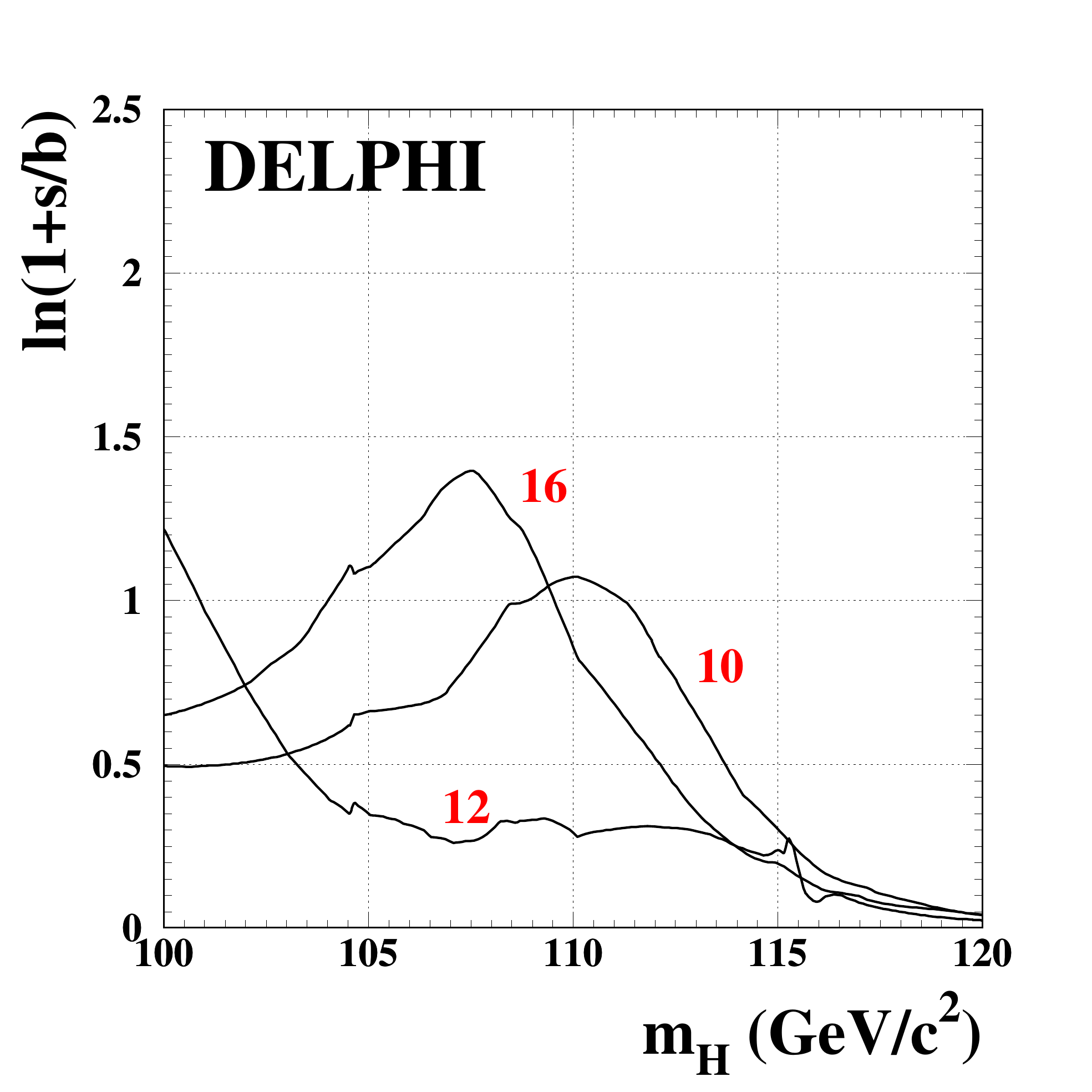}

\includegraphics[width=0.35\linewidth]{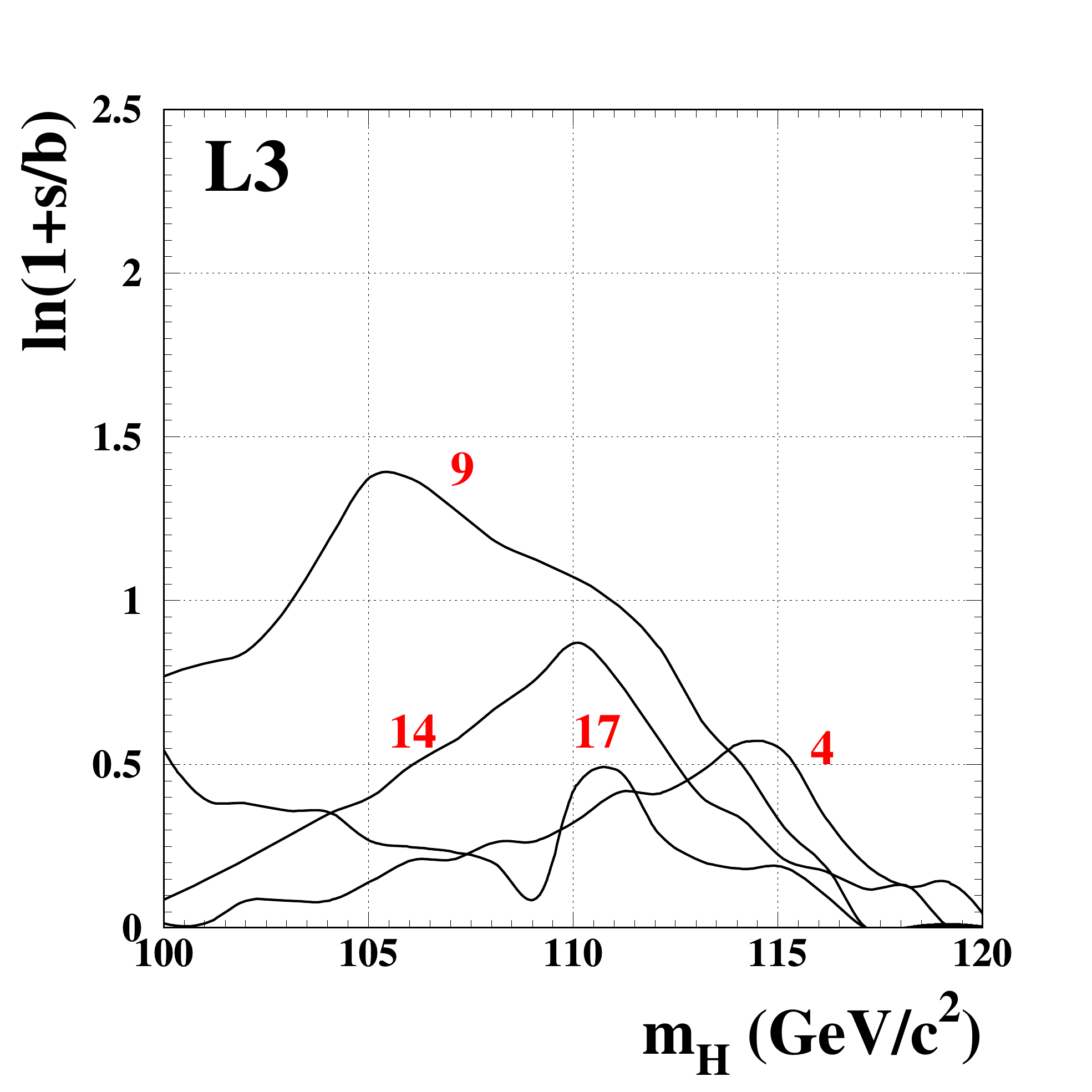} 
\includegraphics[width=0.35\linewidth]{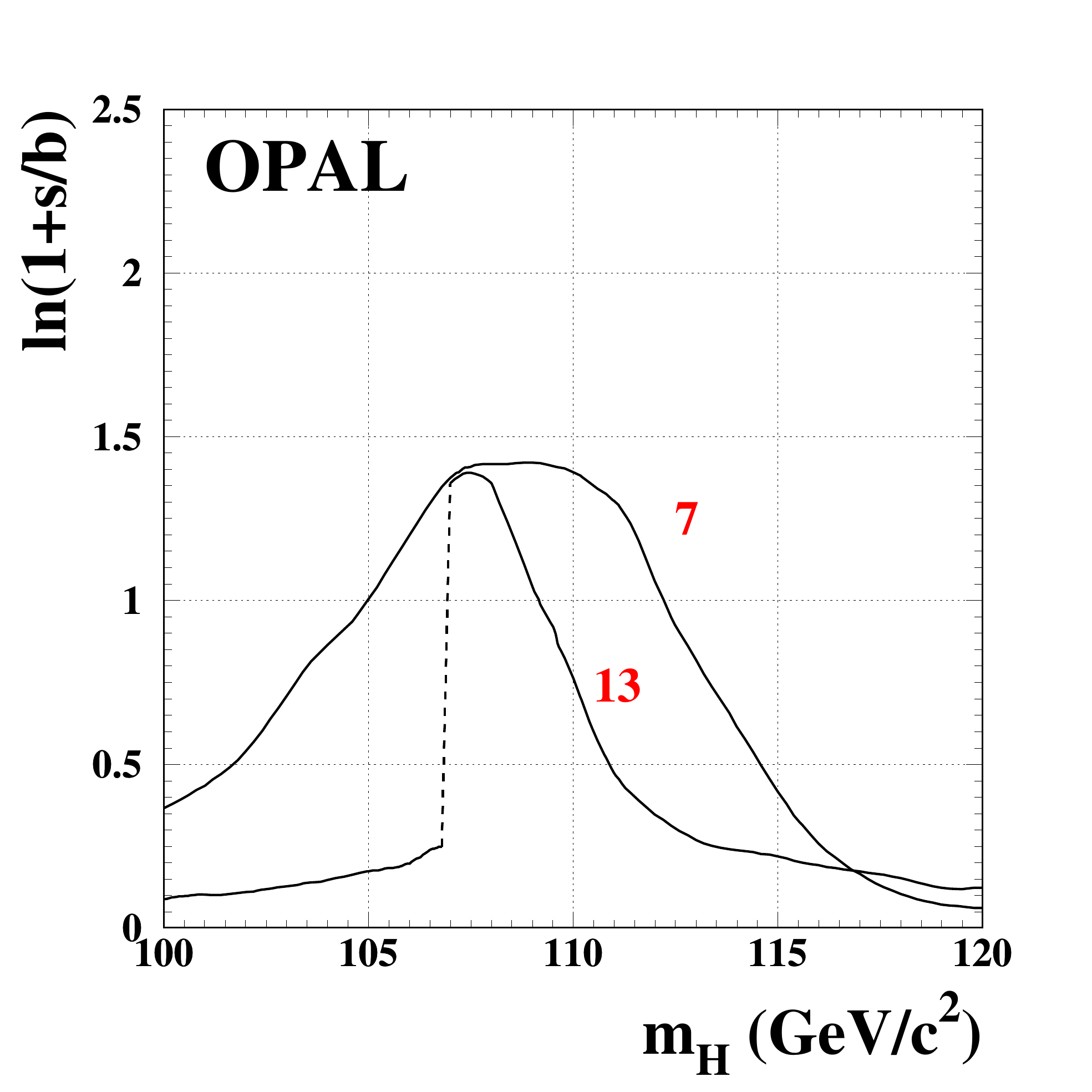}
\end{center}
\caption{\label{fig:spaghetti}{\em Spaghetti diagrams} of 17 Higgs-like events
  detected by the 4 LEP experiments: signal weights against the simulated
  Higgs mass \cite{ref:lephiggs2003}. The ALEPH events crowd around
  115~GeV/$c^2$ whereas for the other 3 experiments there are less of them
  with a rather random mass distribution.}
\end{figure}

A quite interesting feature of data analysis was the plotting of {\em
  spaghetti diagrams}. Those are signal weight distributions of each selected
event as a function of the assumed Higgs mass. Fig.~\ref{fig:spaghetti} shows
the weight distributions of 17 selected Higgs-like candidate events observed
by the 4 LEP experiments \cite{ref:lephiggs2003}. The ALEPH events crowd around
115~GeV/$c^2$ whereas for the other 3 experiments there are less of them with
a rather random mass distribution. This caused quite an excitement at LEP:
many physicists signed the petition to the Director General of CERN to extend
the life of LEP by another year, but that was refused: the simulated
projections were not very promising for a discovery of the SM Higgs boson (the
effect seen by ALEPH only was far too large, much higher than the prediction
of the Standard Model), and the contractors for building LHC were already
prepared to start.

\section{Search and observation at LHC}

\begin{figure}[h]
\begin{center}
\begin{minipage}{0.80\linewidth}
\includegraphics[width=\linewidth]{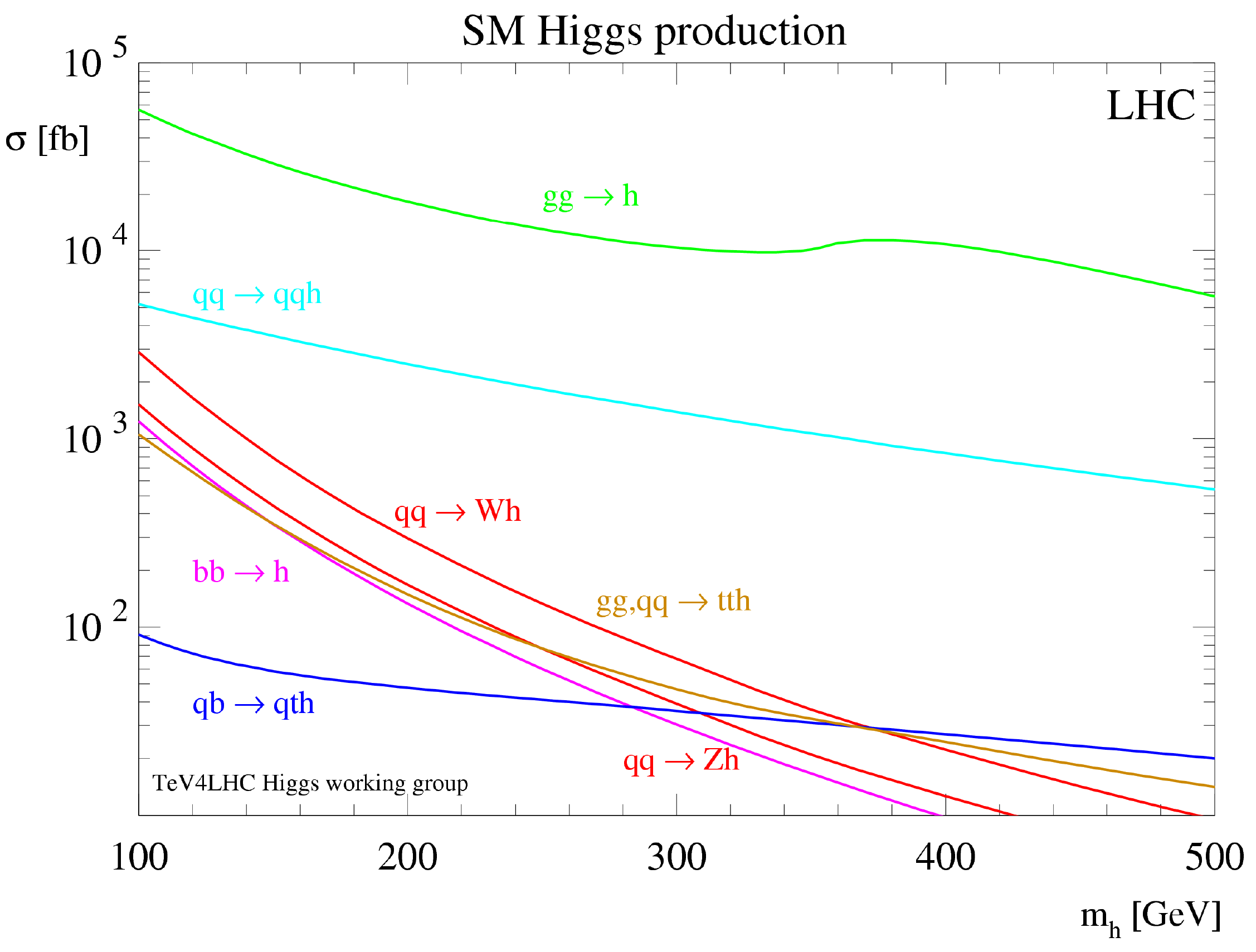}
\end{minipage}
\begin{minipage}{0.15\linewidth}
\begin{center}
\includegraphics[width=\linewidth]{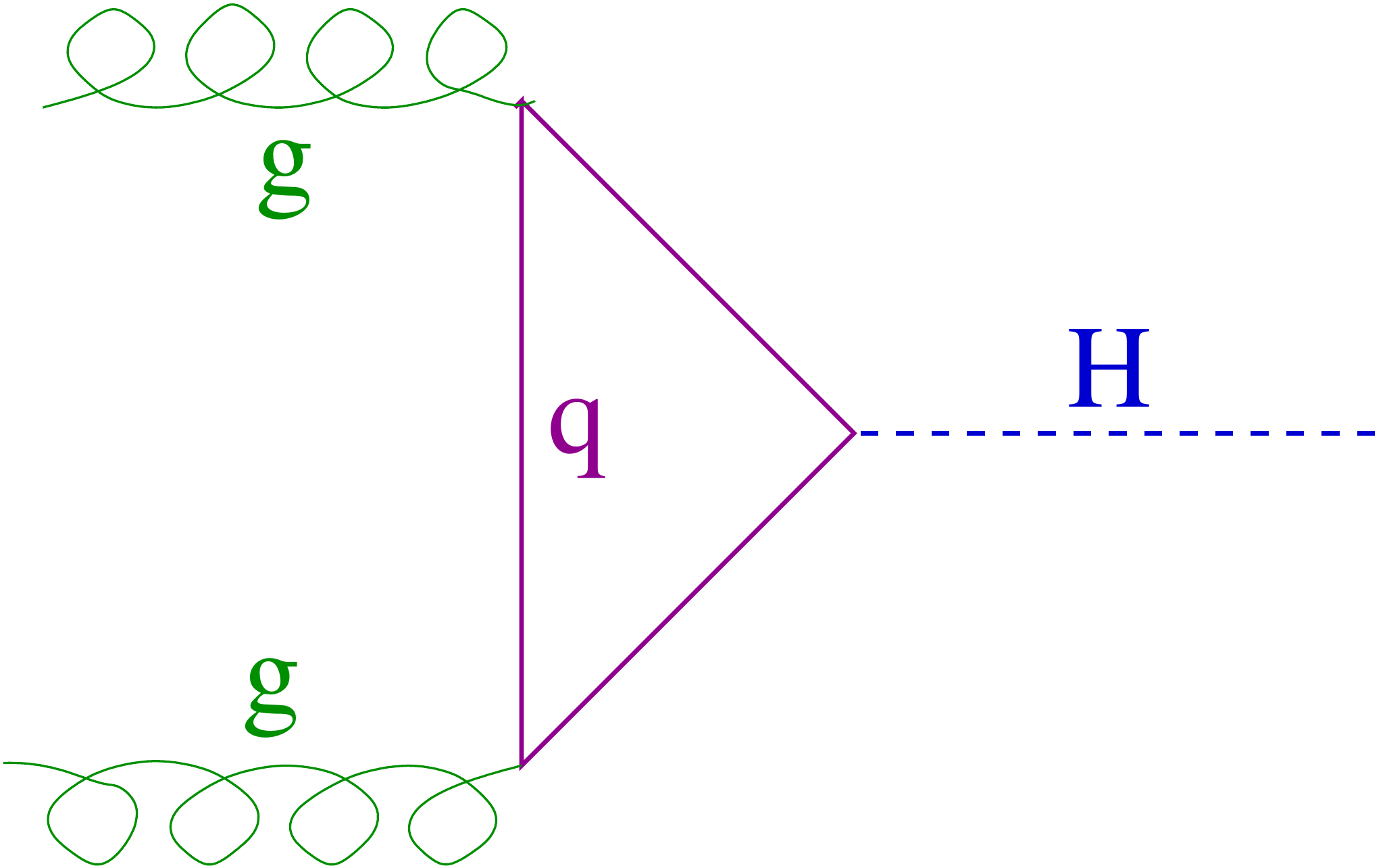}

{\footnotesize
{gluon fusion}

\bigskip\bigskip\bigskip

\includegraphics[width=\linewidth]{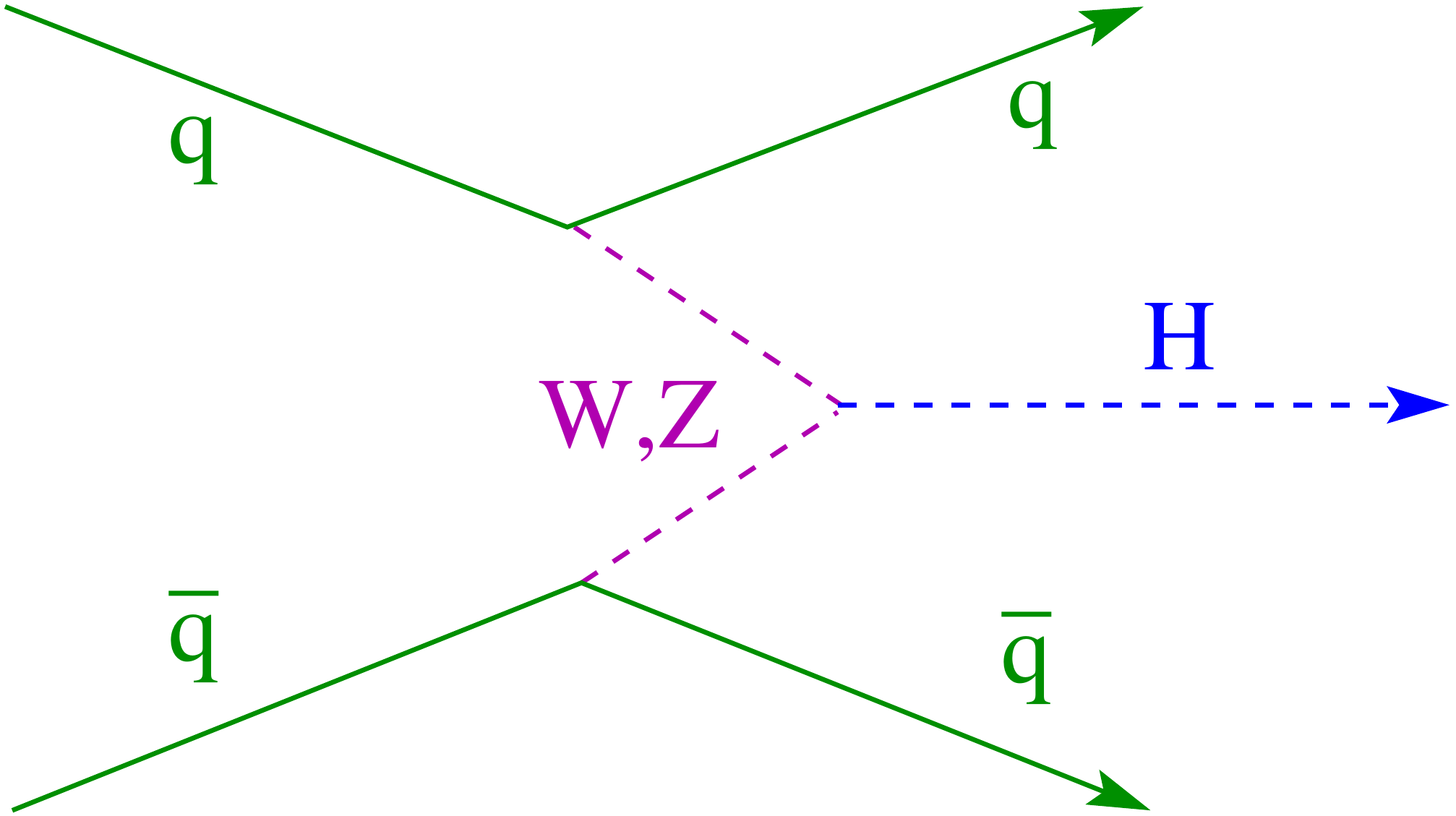}

vector~boson\\ fusion}
\end{center}
\end{minipage}
\end{center}
\caption{\label{fig:SMH-prod}Formation of the SM Higgs boson in p-p collisions
  at LHC.
}
\end{figure}

Figure \ref{fig:SMH-prod} shows the various formation processes of the SM
Higgs boson in p-p collisions at LHC. The dominant reaction is gluon fusion
and vector boson fusion is also significant. 

\begin{figure}[h] 
\begin{center}
\vspace*{-10mm}
\includegraphics[width=0.9\linewidth]{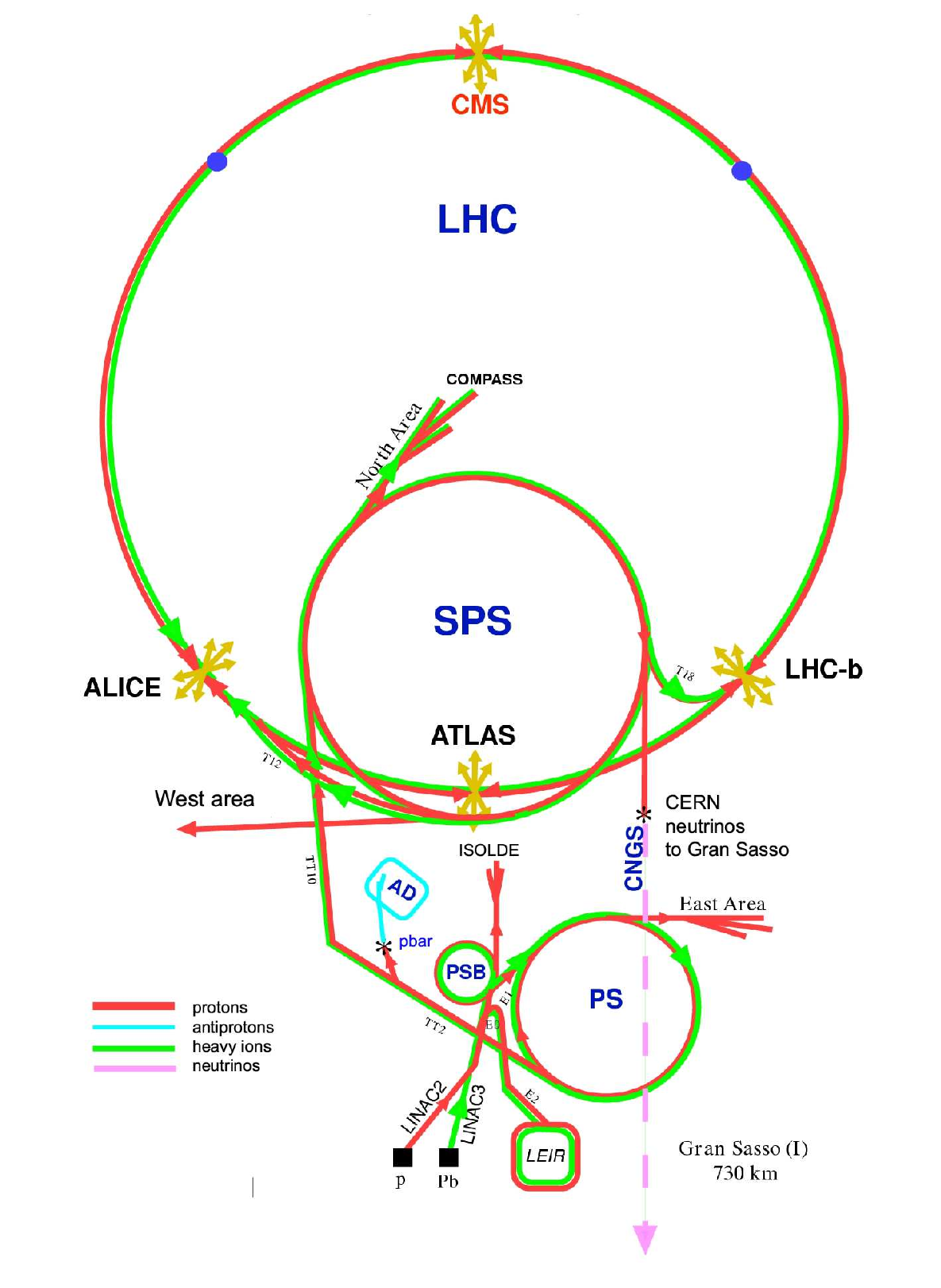}
\caption{The accelerator complex of CERN with the Large Hadron Collider}
\label{fig:CERNAcc05}
\end{center}
\end{figure}

Just like LEP had, the Large Hadron Collider has also 4 interaction points
(Fig.~\ref{fig:CERNAcc05}) with a major experiment (and sometimes a smaller
one as well) in each. The two largest ones, ATLAS \cite{ref:ATLAS} and CMS
\cite{ref:CMS} were designed with the main aim of discovering the Higgs boson,
ALICE \cite{ref:ALICE} is specialized on heavy ion collisions and LHCb
\cite{ref:LHCb} on studying rare processes involving b quarks. The author
belongs to CMS, so most of the results we mention are due to CMS, but all will
be compared to those of ATLAS pointing out the similarities and the (very few
and not significant) differences.  

The ATLAS and CMS collaborations are really huge. According to the official
statistics in 2012 CMS had 3275 physicists (incl. 1535 students) and 790
engineers and technicians from 179 institutions of 41 countries (ATLAS was
even slightly larger). The biggest participant of CMS is the USA, then Italy,
Germany and Russia. It is quite remarkable how similar and different are ATLAS
and CMS. ATLAS uses a lot of new detector techniques while CMS consists of
mostly traditional parts. CMS is based on the largest superconducting solenoid
on Earth whereas ATLAS has a smaller solenoid encircled by 8 huge magnets
making a toroidal field. CMS weighs 14000 tons, twice the weight of ATLAS in
an order of magnitude smaller volume. And in spite of all these differences,
the two collaborations get very similar results.

The design of LHC and its experiments started well before the actual start
of LEP, which means that the construction of the LHC detectors took two 
decades of hard work before the actual data acquisition started. Its first
two years LHC devoted to development rather than data taking, that really
started in 2011 only. 

Even before LHC started the parameter fitting of the Standard Model pointed
toward a light Higgs boson, with a mass around 100~GeV/$c^2$. As LEP excluded
the Higgs boson below 114~GeV/$c^2$ the LHC experiments had to be prepared for
detecting the Higgs boson in the most complicated mass region, around
120~GeV/$c^2$, with several competing decay channels
(Fig.~\ref{fig:SMH_br1}). It was shown very early that the best channels to
observe a light Higgs boson at LHC should be two-photon, $\mrm{H} \ra
\gamma\gamma$ and 4-lepton, $\mrm{H \ra ZZ^* \ra \ell^+\ell^-\ell^+\ell^-}$
channels, because of the very high hadron background. Indeed, in 2012 the LHC
luminosity was already so high that every bunch crossing (ie.\ every event)
contained 10--20 p-p collisions leading to copious hadron production. Thus
both large experiments, CMS and ATLAS designed their tracking systems and
electromagnetic calorimeters with this in mind. The electromagnetic
calorimeter of CMS consists of 75,848 PbWO$_4$ single crystal scintillators,
whereas that of ATLAS is a sampling calorimeter based on liquid argon shower
detectors.

\begin{figure}[h]
\begin{center}
\includegraphics[width=0.7\linewidth]{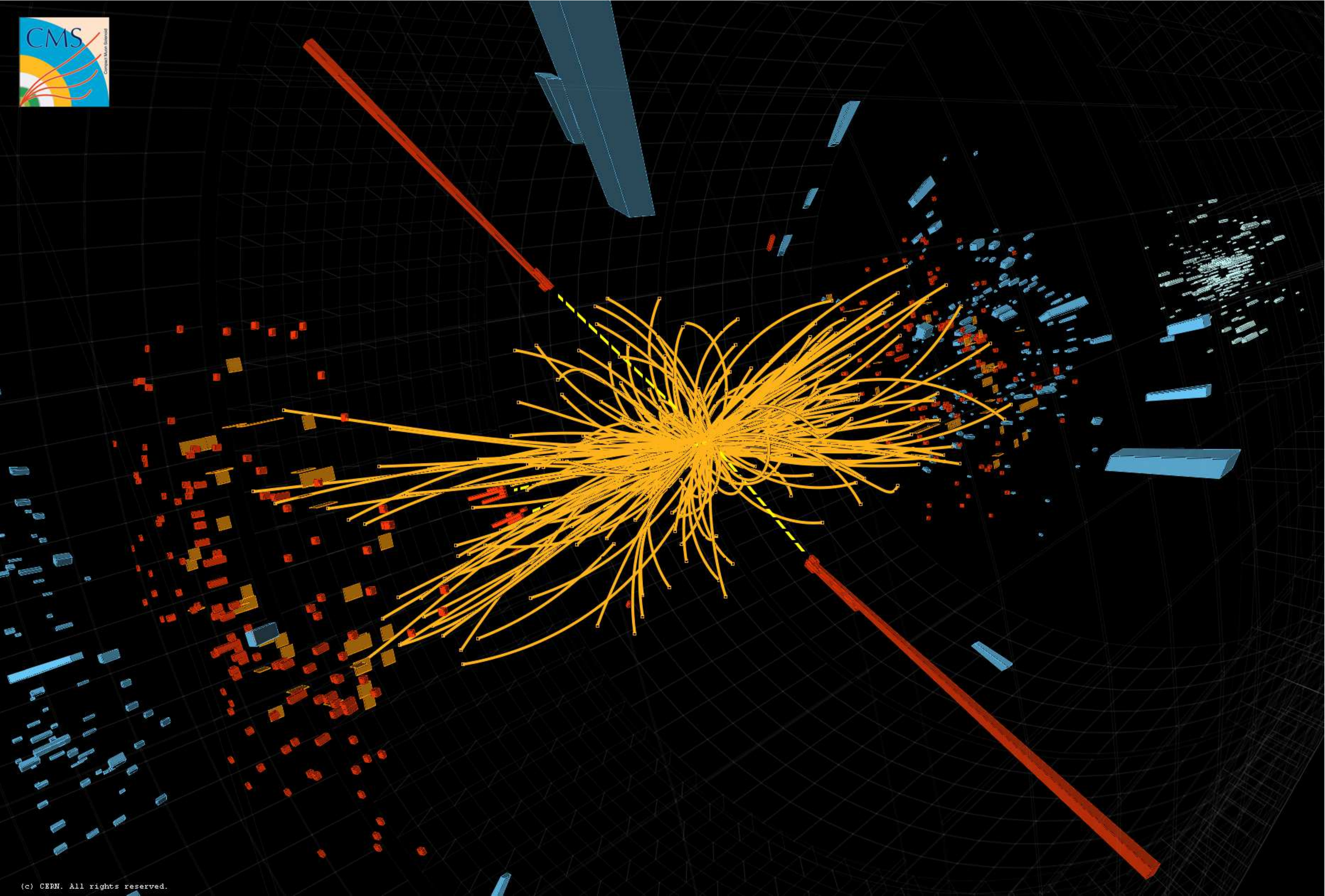}
\caption{\label{fig:cms-gg-event}Higgs-like event detected by the CMS
  experiment at LHC: a boson is formed in a p-p collision and decays to 2 
  energetic gamma photons. The block sizes at the end of the invisible photon
  trajectories correspond to the photon energies deposited in the
  electromagnetic calorimeter.}
\end{center}
\end{figure}

\begin{figure}[h]
\begin{center}
\includegraphics[width=0.8\linewidth]{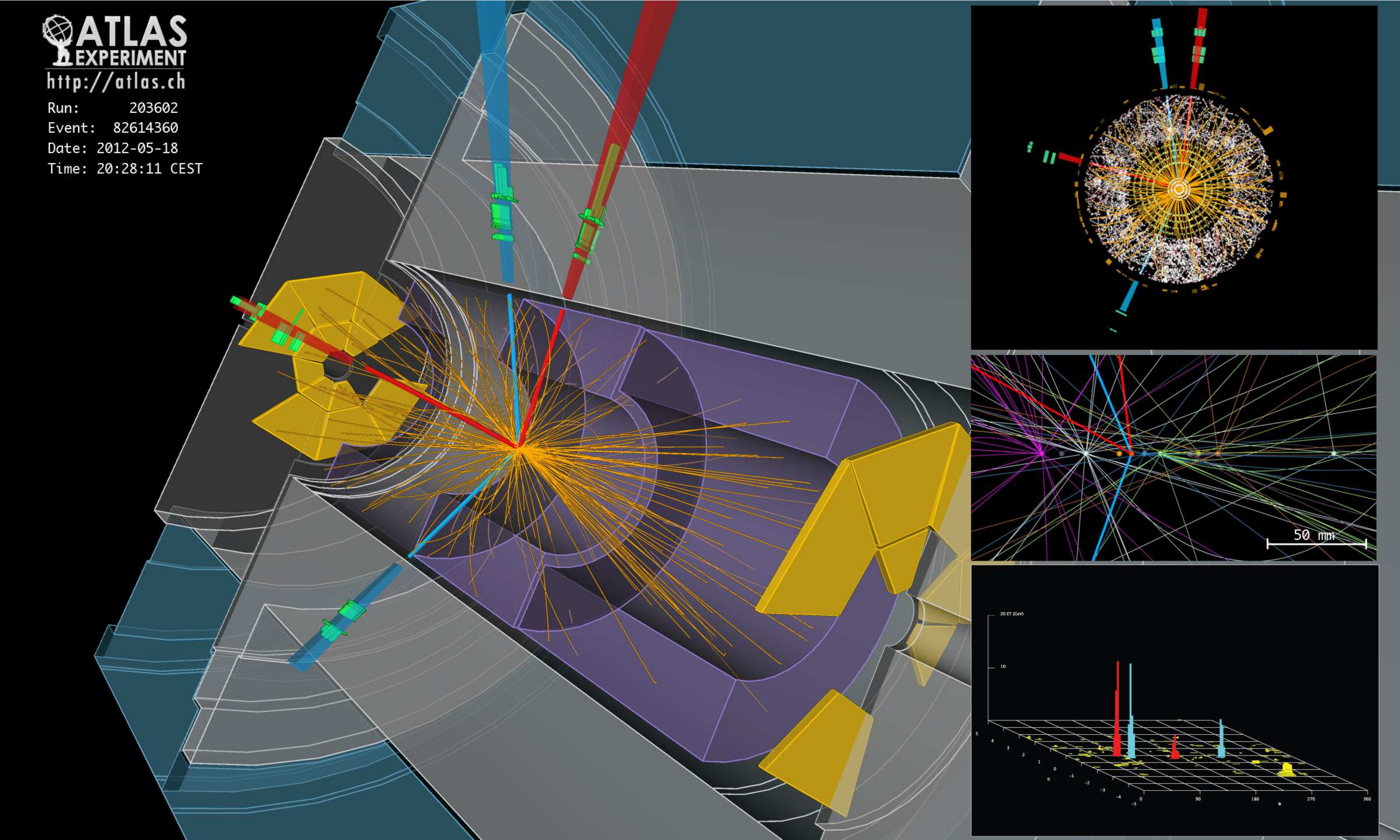}
\caption{\label{fig:atlas-4e-event}Higgs-like event detected by the ATLAS
  experiment at LHC: a p-p collision produces 4 electrons, one pair is from
  Z-decay as identified by the invariant mass. The 4-electron mass corresponds
  to a decaying mass of 125 GeV/$c^2$. Upper right: x-y view, lower right:
  lego plot of energy deposits in the calorimeters. Middle right shows that
  there were 9 other identified vertices of p-p collisions in the
  neighborhood of the candidate Higgs decay belonging to the same LHC bunch
  crossing.  }
\end{center}
\end{figure}

By the beginning of 2012, when all 2011 data were analyzed, the possible mass
of the SM Higgs boson was already confined to the region of 114 $< M_\mrm{H}<$
127 GeV/$c^2$ by CMS~\cite{ref:CMS_H-comb_2011} (with very similar results
from ATLAS). In that region 2-- 3 $\sigma$ excesses were found at
$\sim$125 GeV/$c^2$ in the two main decay channels, H~\ra$\gamma\gamma$ and
H~\ra~ZZ. It seemed more and more probable that
the Higgs boson will be observed at LHC in 2012, it was even decided by the
CERN administration to extend the data taking scheduled for 2012 before the
long shutdown for accelerator development if necessary for the discovery.

On July 4th, the beginning of the large annual high-energy physics congress in
Melbourne, the spokespersons of ATLAS and CMS gave talks from CERN (in
internet connection to the whole word, including, of course, the main
auditorium of the Australian conference) on Higgs search. They announced that
at LHC collision energies 7 and 8 TeV, in the two most significant decay
channels H~\ra~$\gamma\gamma$ and H~\ra~ZZ~\ra~$\ell^+\ell^-\ell^+\ell^-$, at
an invariant mass of {$m\approx 126$ GeV} a new boson is seen at a convincing
statistical significance of $5\sigma$ confidence level each with properties
corresponding to those of the Standard Model Higgs boson. The fact that the
new particle could decay to two photons or Z bosons, confined its spin to an
even integer, i.e. a boson of $S=0$ or $S=2$. Of course, as the data analysis
was optimized to find the SM Higgs, it was very unlikely to find something
very different. Nevertheless, the two experiments emphasized that it has to be
studied, whether or not its spin is really zero with a + parity (the
pseudo-scalar mesons have spin 0 with negative parity), and that its decay
probabilities to various final states follow the predictions of the Standard
Model. After reanalyzing their data the Tevatron experiments, CDF and D0 also
found an excess \cite{ref:cdf-d0_2012} at this mass (after the LHC started the
Tevatron accelerator of Fermilab was stopped).

\section{Reactions of the Media}

 The saying that {\em three people can keep something secret only if two of
   them are dead} is attributed to Benjamin Franklin. As any result of a
 collaboration has to be approved by all members before it is made public, the
 more than 6000 participants of ATLAS and CMS knew well in advance the
 developing result. Thus two days before the 4th July announcement, {\em
   Nature Online} already reported the result~\cite{ref:Nature2July}.  Of
 course, the fact that the CERN management invited to the seminar all leading
 scientists of the field including the theoreticians who developed spontaneous
 symmetry breaking for the Standard Model also helped people to guess that
 something dramatic will be announced. 

 CERN produced some figures concerning the media echo of the day: 55 media
 organizations were represented at the talks of 4 July, the talks were
 broadcasted via close to half a million internet connections (many of them
 being conference rooms in partner institutions, e.g. three in Hungary with
 quite an audience in each), 1034 TV stations devoted 5016 news broadcasts to
 the event for more than a billion ($10^9$) people. Many-many news articles
 and even more blogs and talks discussed the conditions and importance of the
 discovery.

\section{The observations}

\begin{figure}[h]
\begin{center}
\includegraphics[width=0.49\linewidth]{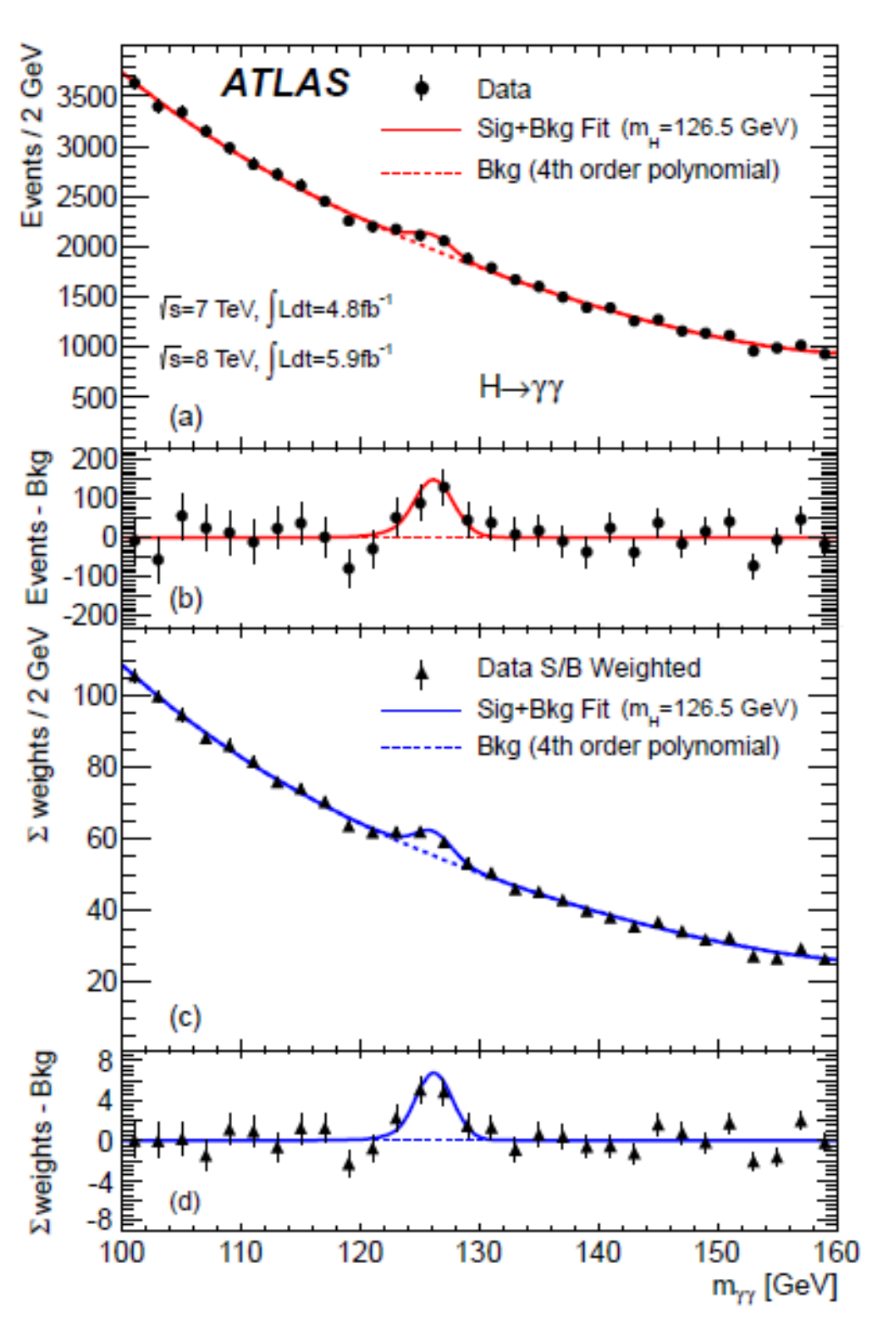}
\includegraphics[width=0.49\linewidth]{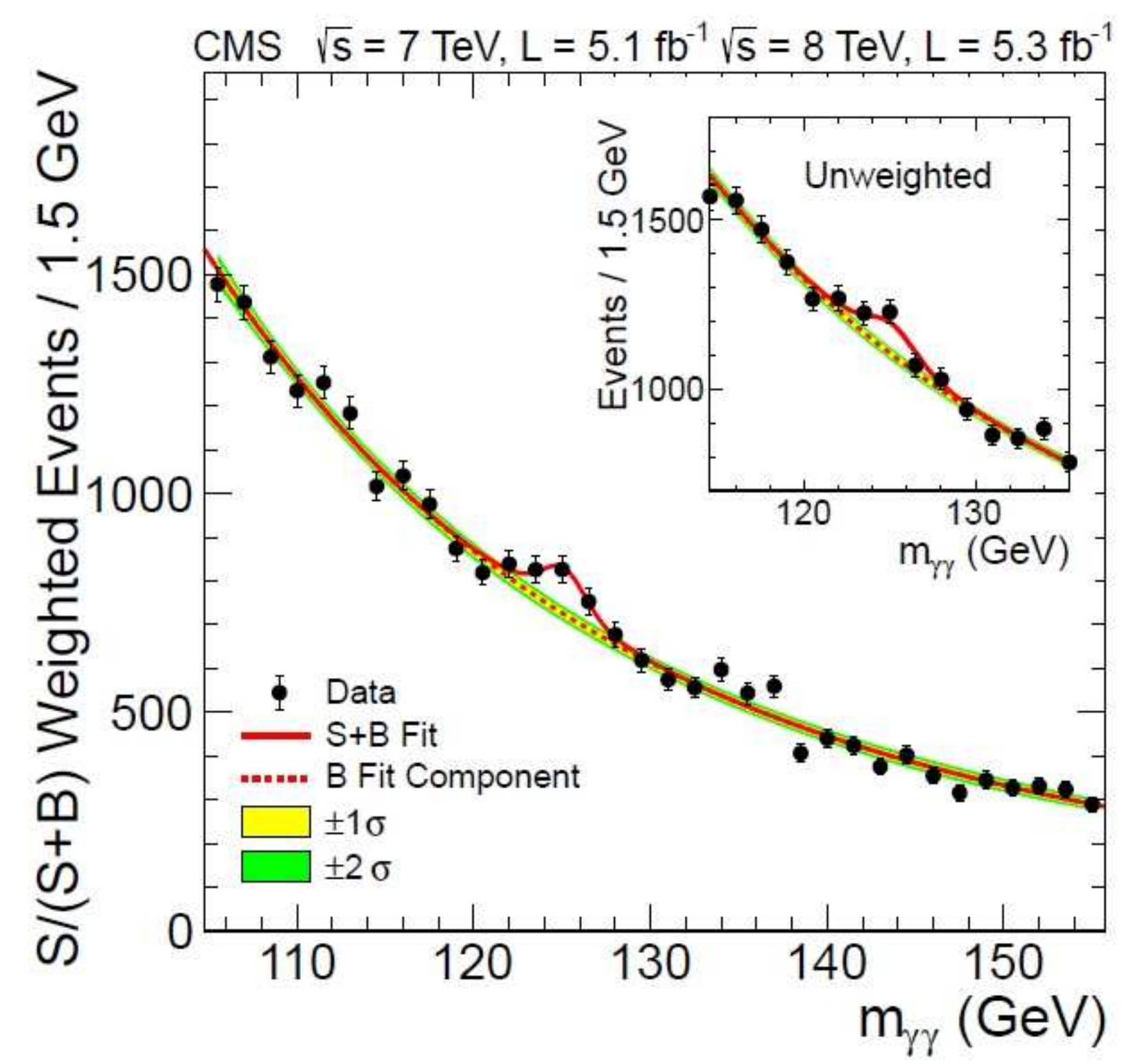}
\end{center}
\caption{\label{fig:h2gg_07.12}First observation of the Higgs-like boson by
  ATLAS \cite{ref:atlas_higgs_20120817} and CMS \cite{ref:cms_higgs_20120817}
  in the $\gamma\gamma$ invariant mass distribution at 126 GeV/$c^2$. The
  amplitudes of both observed signals are close to the expectations of the
  Standard Model as shown by the fitted curves. Both experiments plotted the
  raw events and also the sum of event weights according to their
  signal-likelihood.  }
\end{figure}

\begin{figure}[h]
\begin{center}
\includegraphics[width=0.49\linewidth]{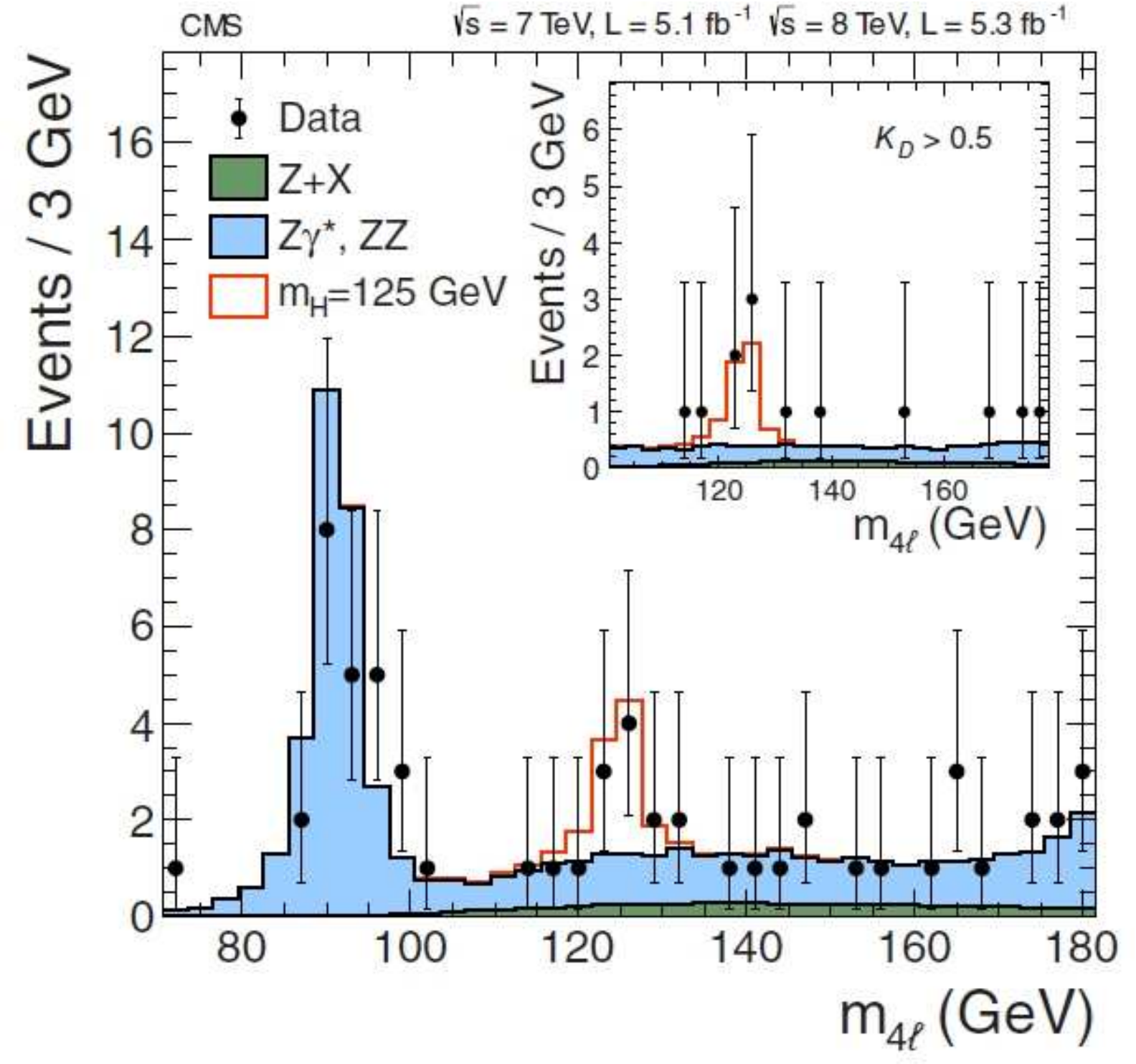}
\includegraphics[width=0.49\linewidth]{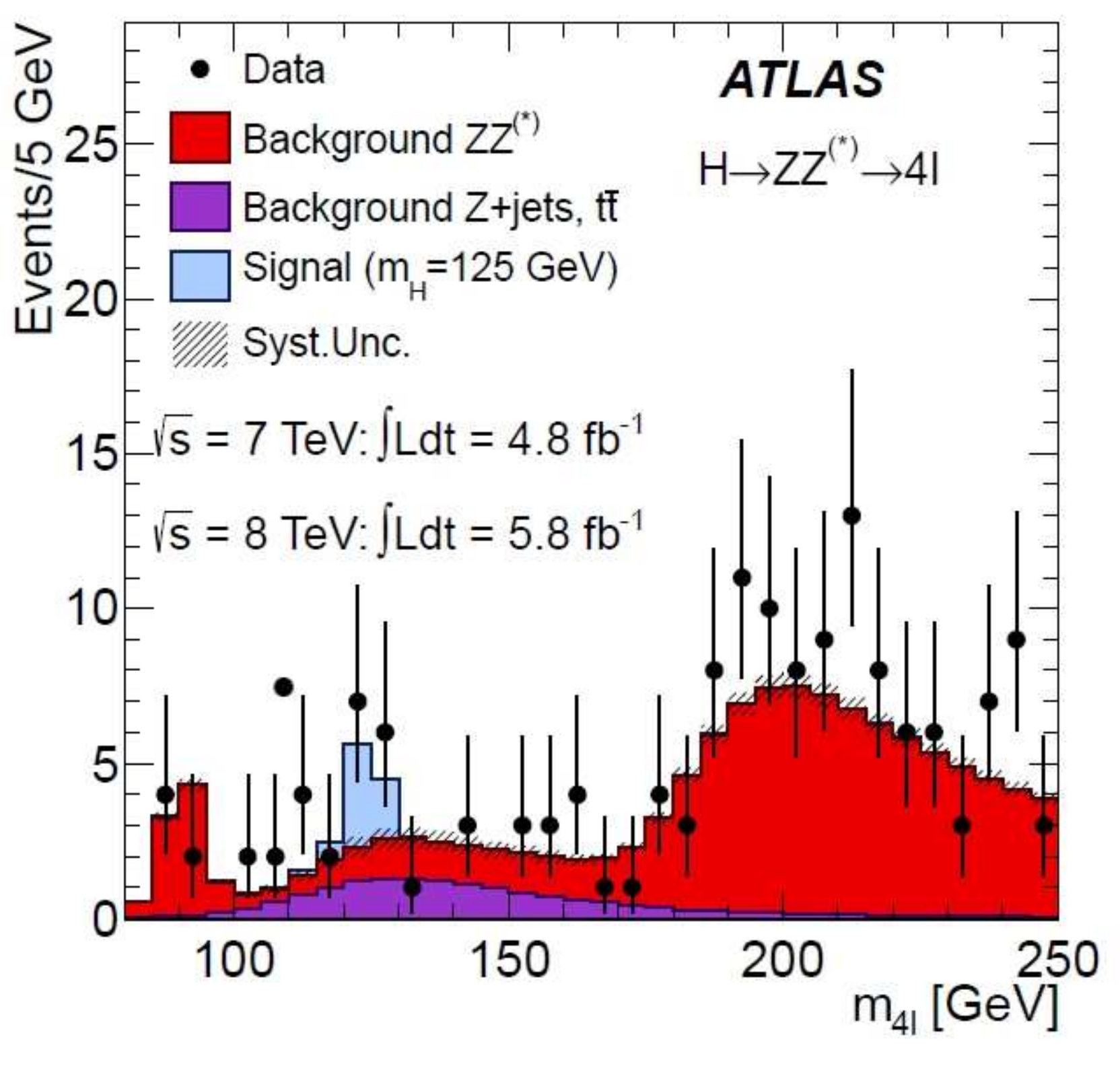}
\end{center}
\caption{\label{fig:h2llll_07.12}First observation of the Higgs-like boson by
  CMS \cite{ref:cms_higgs_20120817} and ATLAS \cite{ref:atlas_higgs_20120817}
  in the $\ell^+\ell^-\ell^+\ell^-$ invariant mass distribution at 126
  GeV/$c^2$. The amplitudes of the observed signals are close to the
  expectations of the Standard Model.}
\end{figure}

On 31 July the two experiments submitted papers of the discovery to Physics
Letters B, they were published 14 August \cite{ref:atlas_higgs_20120817,
  ref:cms_higgs_20120817}. Both papers are 15 pages long followed by 16 pages
of close to 3000 authors and both are dedicated to the memory of those
participants who could not live to see the result of the more than two decades
of construction work. Fig.~\ref{fig:h2gg_07.12} shows the di-photon spectra
obtained by ATLAS and CMS in July 2012, after analyzing about a quarter of the
data to be collected in 2012. The 4-lepton spectra were quite similar with
less background and signal, they are shown in Fig.~\ref{fig:h2llll_07.12}.
Thus both experiments saw at the same invariant mass in both most significant 
decay channels the new boson. In all cases the signal strengths agreed within
uncertainties to the predictions of the Standard Model.

\begin{figure}[h]
\begin{center}
\includegraphics[width=0.9\linewidth]{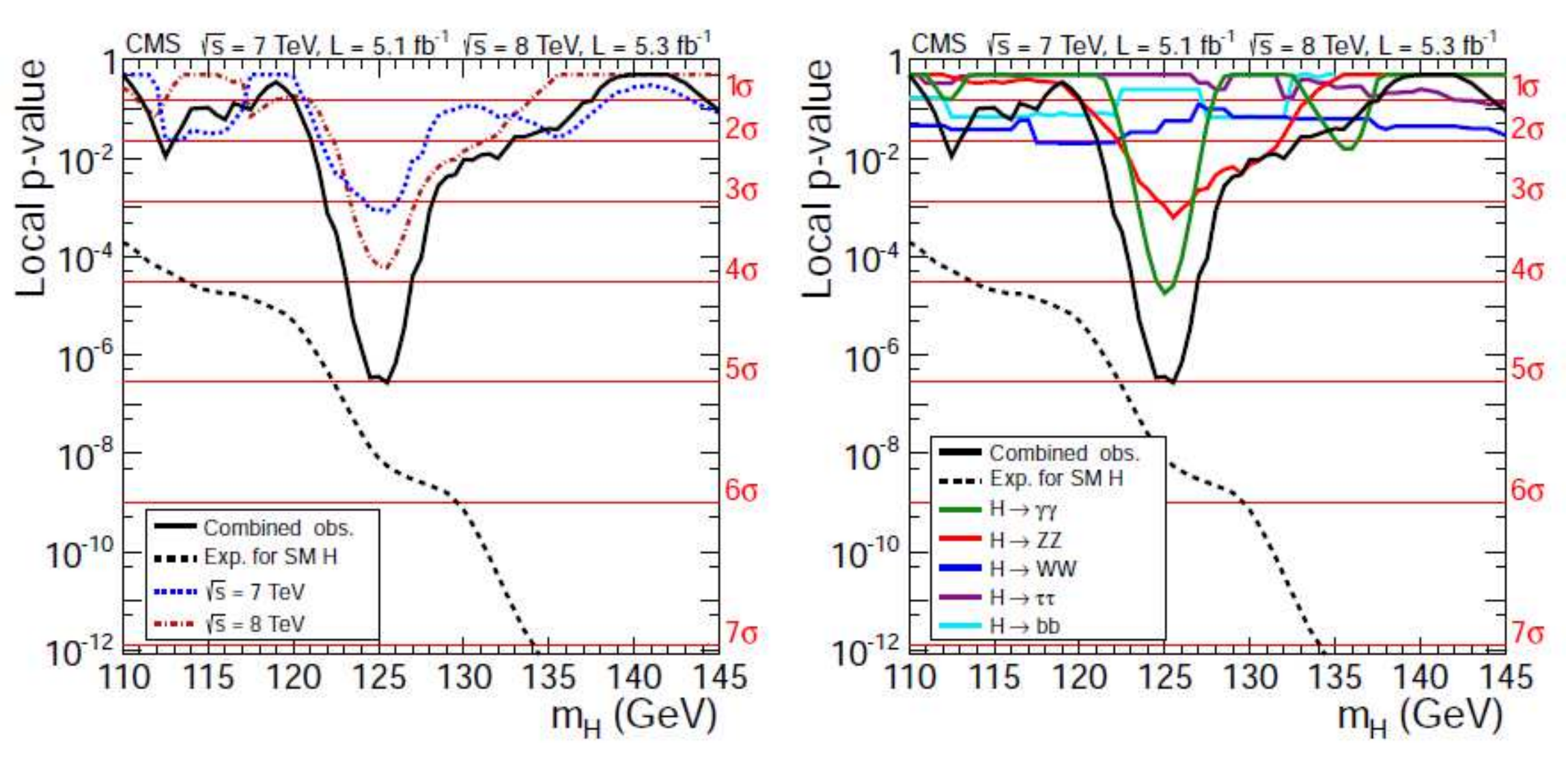}
\end{center}
\caption{\label{fig:cms_p-val}Observation of the Higgs-like boson by CMS
  \cite{ref:cms_jhep_obs} in the invariant mass distribution of p-values at
  125 GeV/$c^2$ as based on the data collected at LHC at collision energies 7
  TeV in 2011 and 8 TeV in 2012 (left) and in the various decay channels
  (right). The amplitude of the observed signal is close to the expectations
  of the Standard Model.}
\end{figure}

What was really convincing of the observation was the distribution of the
p-values of the events selected in the various analyzed decay channels of the
hypothetical Higgs boson. For CMS it is shown in Fig.~\ref{fig:cms_p-val}: the
significance is already as high as $5\sigma$ for the data available in July
2012.  It was a joke of statistics that in July 2012 adding together two decay
channels, H~\ra~$\gamma\gamma$ and H~\ra~$4\ell$ gave the same $5\sigma$
significance for both ATLAS and CMS whereas adding to it the results for other
channels increased the significance to $6\sigma$ for ATLAS and left it at
$5\sigma$ for CMS (Fig.~\ref{fig:cms_p-val}).

\section{Is it really the Higgs boson?}

\begin{figure}[h]
\begin{center}
\includegraphics[width=0.6\linewidth]{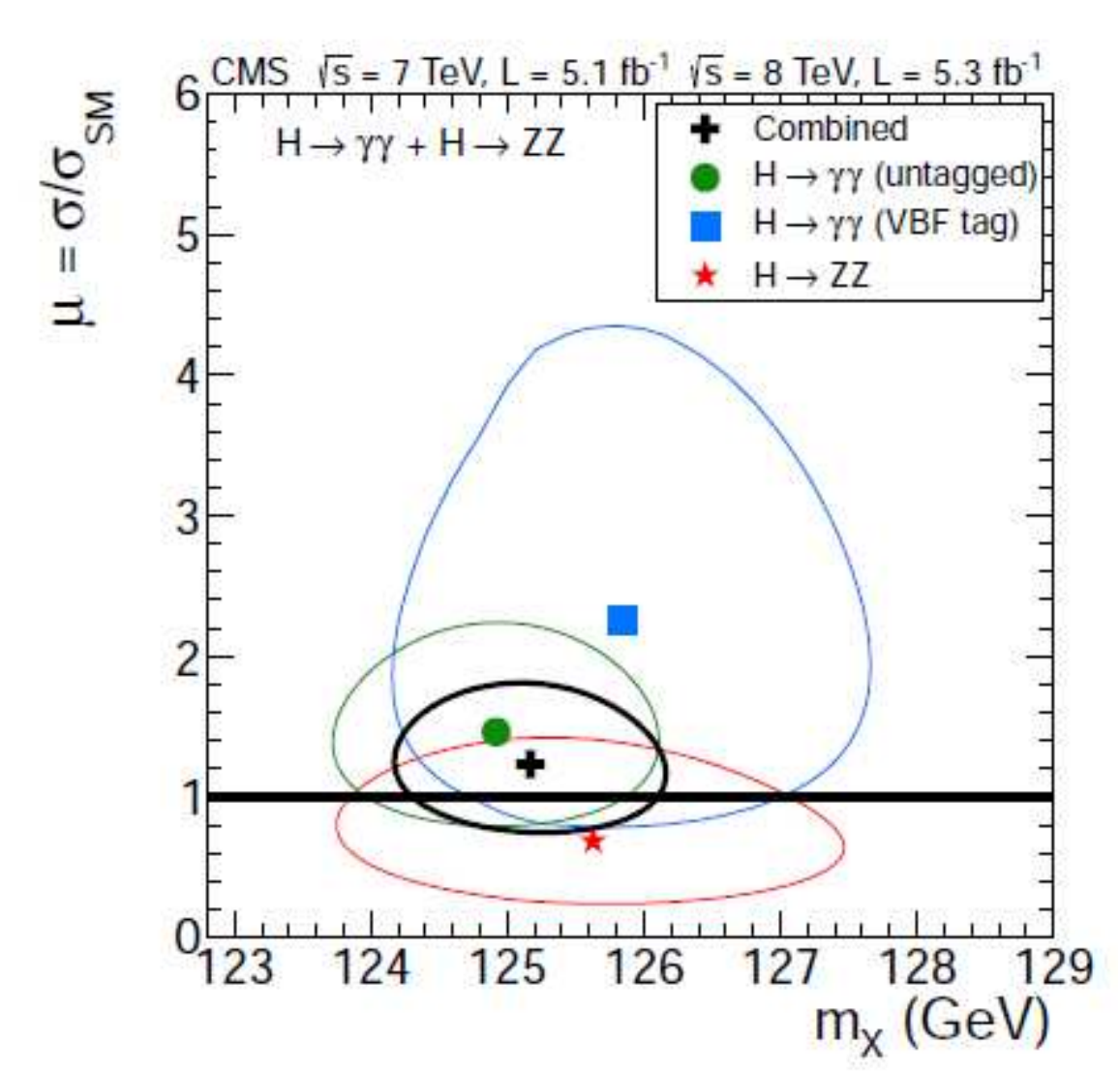}
\end{center}
\caption{\label{fig:cms_mass_xsec}The relative signal strengths in the most
  significant decay channels normalized to the predictions of the Standard
  Model for $M_\mrm{H}=125$~GeV against the invariant masses of the new boson
  as measured by CMS\cite{ref:cms_jhep_obs}. The two data points agree within
  statistics with each other and with the expectations of the Standard Model.}
\end{figure}

Analyzing most of the data collected in 2012 confirmed the existence of the
new boson and led to the conclusion that all observed properties of the newly
discovered particle are within statistics close to those predicted for the
Higgs boson of the Standard Model. The fact that it decays to two photons
points to its having spin 0 or 2. The charged lepton spectra bears the
features of its having $S=0^+$ as ascertained by both experiments
\cite{ref:atlas_2013_spin,ref:cms_2013_spin}. Its mass as determined by
CMS~\cite{ref:Bluj2013}, the average of all decay channels is $<M_\mrm{X}> =
\mrm{125.7 \pm 0.3 (stat) \pm 0.3 (syst)}$ (Fig.~\ref{fig:cms_mass_xsec}). The
ATLAS result is almost exactly the same \cite{ref:Bluj2013}: $\mrm{125.5 \pm
  0.2 (stat) \pm0.6 (syst)}$. The differences in the uncertainties are due to
the facts that ATLAS had more signal-like data, but also got more different
masses in the two main channels. The measured signal strengths of the new
particle are also compatible with that expected for the Standard Model Higgs
boson: for CMS it is $\sim20\%$ less while for ATLAS $\sim40\%$ more than the
SM prediction, but both deviations are within the experimental
uncertainties. As a theoretician remarked whenever ATLAS has an excess CMS
comes up for everybody's annoyance with a deficit, bringing the average close
to the SM prediction.

\begin{figure}[h]
\begin{center}
\includegraphics[width=0.9\linewidth]{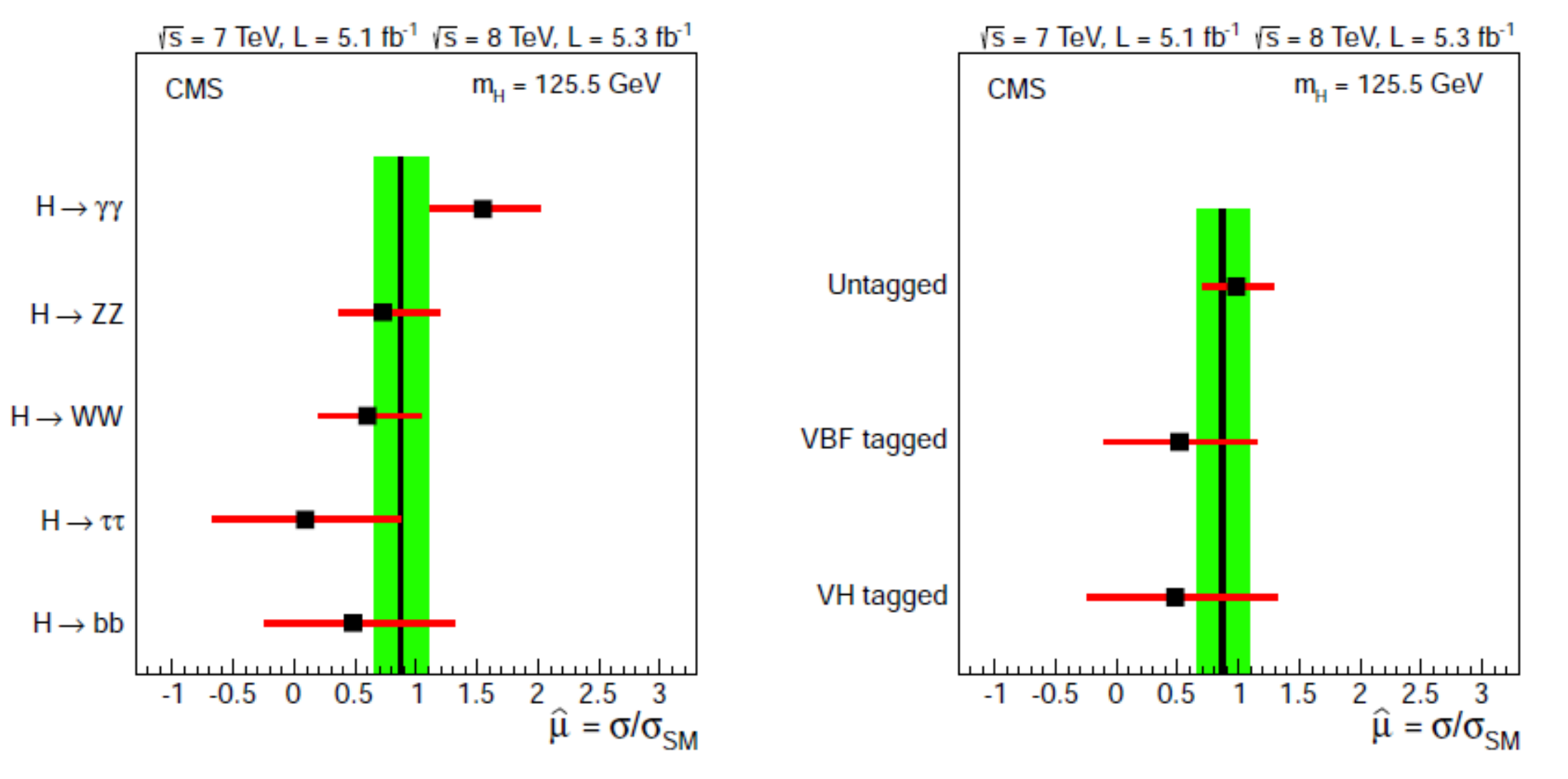}
\end{center}
\caption{\label{fig:cms_sig-str}Signal strengths observed by CMS
  \cite{ref:cms_jhep_obs} for the Higgs-like particle in various possible
  decay (left) and production (right) channels as compared to those predicted
  by the Standard Model for the Higgs boson with a mass of 125.5 GeV/$c^2$.}
\end{figure}

The LHC experiments studied the cross sections of the processes connected to
the new particle. Fig.~\ref{fig:cms_sig-str} shows the signal strengths of
production and decay in various possible channels of the Higgs-like boson
measured by CMS \cite{ref:cms_jhep_obs} as compared to those predicted by
the Standard Model for the Higgs boson with a mass of 126 GeV/$c^2$.  The
amplitudes of all observed signals are in agreement with the expectations of
the Standard Model. ATLAS got similar results, of course. 

Thus what we found is very likely the Standard Model Higgs boson. On one hand
this is a great success of particle physics. On the other hand this is
somewhat of a disappointment as the SM has theoretical shortcomings which need
new physics to resolve. Just to list a few of them: it cannot unite the
interactions at large energies, cannot account for the dark matter of the
Universe and cannot explain neutrino oscillations. There are many extensions
of the theory which should result in deviations from the Standard Model. Most
of those problems can be resolved e.g.\ by {\em supersymmetry}, but none of
its predicted phenomena could be found yet experimentally. The observables of
the Higgs boson should be sensitive to some of the features of new physics and
these studies will be the main job of ATLAS and CMS in the future, from 2015
when the LHC will restart with twice the energy and luminosity of 2012.

It is very interesting that the 126 GeV mass of the Higgs boson seems to be
exciting for theoreticians, there was even a special workshop
\cite{ref:whm126} organized to discuss this mass in 2013. The reason is that
$M_H=126$~GeV is at the border line of the stability of electroweak vacuum on
the plane of top mass against Higgs mass, see e.g.~\cite{ref:alekhin12}. At
the Madrid workshop the apparent fine tuning of the Standard Model compelled
some physicists to recall the anthropic principle.

\section{The Nobel Prize}
 
As the LHC discovery gradually ripened many people asked who will get the
Nobel Prize if the Higgs boson is discovered. There were even rumors that the
Nobel Committee might consider the give it to CERN and to the two
collaborations, although that would have been against the Nobel tradition. By
March 2013 it was already quite clear that the new particle is actually a
Higgs boson and very probably that of the Standard Model. In the end of
September 2013 the blogs started speculate about Higgs and Englert as Robert
Brout passed away in 2011. Indeed, in October Fran\c{c}ois Englert and Peter
Ware Higgs received the 2013 Nobel Prize in physics "for the theoretical
discovery of a mechanism that contributes to our understanding of the origin
of mass of subatomic particles, and which recently was confirmed through the
discovery of the predicted fundamental particle by the ATLAS and CMS
experiments at CERN's Large Hadron Collider."

\section{Acknowledgments}
The results reported here are due to the Higgs groups of LEP and of the LHC
Collaborations. The author is indebted to Tommaso Dorigo, Paris Sphicas and
Michael Spira for useful suggestions. This work was supported by the Hungarian
National Science Foundation OTKA via contracts NK-81447, K-103917 and K109703.

\flushleft


\begin{thebibliography}{99}
\bibitem{ref:Englert} Englert F and Brout R, 
``Broken Symmetry and the Mass of Gauge Vector Mesons,''
{\it Phys.Rev.Lett.} 13 (1964) 321-323;\\ 
\bibitem{ref:HiggsA}Higgs P W, 
``Broken Symmetries, Massless Particles and Gauge Fields,'' 
{\it Phys.Lett.} 12 (1964) 132-133;\\ 
\bibitem{ref:HiggsB}Higgs P W, 
``Broken Symmetries and the Masses of Gauge Bosons,''
{\it Phys.Rev.Lett.} 13 (1964) 508-509;\\ 
\bibitem{ref:HiggsC}Higgs P W,
``Spontaneous Symmetry Breakdown without Massless Bosons,''
{\it Phys.\ Rev.} {\bf 145} (1966) 1156.\\
\bibitem{ref:Guralnik}Guralnik G S, Hagen C S and Kibble T W B, 
``Global Conservation Laws and Massless Particles,''
{\it Phys.Rev.Lett.} 13 (1964) 585-587.
\bibitem{ref:ewwg} The LEP Electroweak Working Group, 
{\it http://lepewwg.web.cern.ch}
\bibitem{ref:Higgs2002}Higgs P W, 
``My Life as a Boson: The Story of 'The Higgs''', 
{\it Int. J. Mod. Phys. A} {\bf 17} Suppl.~1  (2002) 86-88.
\bibitem{ref:hdstat}Horv\'ath D for the CMS Collaboration, ``Search for the
  Higgs Boson: a Statistical Adventure of Exclusion and Discovery,'' {\it
    arXiv:1310.6839 [hep-ex]}, 2013.
\bibitem{ref:phys-stat-2011}Prosper H B and Lyons L, editors, 
  ``{Proceedings of the PHYSTAT 2011 Workshop on Statistical Issues Related to
  Discovery Claims in Search Experiments and Unfolding}'', CERN, Geneva,
  Switzerland, 17–20 January 2011; CERN-2011-006. 
\bibitem{ref:cous-high}Cousins R D and Highland V L,
  ``Incorporating Systematic Uncertainties into an Upper Limit,''
  {\it Nucl.\ Instrum.\ Meth.\ A} {\bf 320} (1992) 331. 
\bibitem{ref:blind}Klein J R and Roodman A,
``Blind Analysis in Nuclear and Particle Physics,''
  {\it Ann.\ Rev.\ Nucl.\ Part.\ Sci.}  {\bf 55} (2005) 141.
\bibitem{ref:BitKras}Bityukov S I and Krasnikov N V, 2000
``On the Observability of a Signal above Background,''
  {\it Nucl.\ Instrum.\ Meth.\ A} {\bf 452} (2000) 518.
\bibitem{ref:lephiggs2003} The LEP Experiments, ALEPH, DELPHI, L3 and OPAL, 
``Search for the Standard Model Higgs Boson at LEP,''
{\it Phys. Lett. B} {\bf 565} (2003) 61-75.
\bibitem{ref:ATLAS}ATLAS Collaboration, ``A Toroidal Lhc ApparatuS'', 
{\em http://atlas.web.cern.ch/Atlas/Collaboration/}
\bibitem{ref:CMS}CMS Collaboration, ``Compact Muon Solenoid'', 
{\em http://cms.web.cern.ch/}
\bibitem{ref:ALICE}ALICE Collaboration, ``A Large Ion Collider Experiment'', 
{\em http://aliweb.cern.ch/}
\bibitem{ref:LHCb}LHCb Collaboration, {\em http://lhcb.web.cern.ch/lhcb/}
\bibitem{ref:CMS_H-comb_2011}Chatrchyan S {\it et al.}, The CMS Collaboration, 
``Combined Results of Searches for the Standard Model Higgs Boson in 
pp Collisions at $\sqrt{s}=7$ TeV,''
{\it Physics Letters B} {\bf 710} (2012) 26-48.
\bibitem{ref:cdf-d0_2012}Aaltonen T {\it et al.}  [CDF and D0
  Collaborations], ``Evidence for a Particle Produced in Association with Weak
  Bosons and Decaying to a Bottom-Antibottom Quark Pair in Higgs Boson
  Searches at the Tevatron,'' Phys.\ Rev.\ Lett.\ {\bf 109} (2012) 071804
\bibitem{ref:Nature2July} Chalmers M, 
``Physicists Find New Particle, but is it the Higgs?'',
{\em
  http://www.nature.com/news/physicists-find-new-particle-but-is-it-the-higgs-1.10932}, 2 July 2012,
\bibitem{ref:atlas_higgs_20120817} Aad G {\it et al.}, The ATLAS Collaboration, 
``Observation of a New Particle in the Search for the Standard Model 
Higgs Boson with the ATLAS Detector at the LHC,''
{\it Physics Letters B} {\bf 716} (2012) 1--29.
\bibitem{ref:cms_higgs_20120817}Chatrchyan S {\it et al.}, 
The CMS Collaboration, 
``Observation of a New Boson at a Mass of 125 GeV with the CMS Experiment at the LHC,''
{\it Physics Letters B} {\bf 716} (2012) 30--61.
\bibitem{ref:cms_jhep_obs}Chatrchyan S {\it et al.}, The CMS Collaboration,
 ``Observation of a New Boson with Mass near 125 GeV in pp Collisions at $\sqrt{s}$ = 7 and 8 TeV,''
  JHEP {\bf 1306} (2013) 081
\bibitem{ref:atlas_2013_spin}Aad G {\it et al.}, The ATLAS Collaboration,
  ``Evidence for the Spin-0 Nature of the Higgs Boson Using ATLAS Data,''
  Phys.\ Lett.\ B {\bf 726} (2013) 120.
\bibitem{ref:cms_2013_spin}Chatrchyan S {\it et al.}, The CMS Collaboration, 
``Study of the Mass and Spin-Parity of the Higgs Boson Candidate Via Its Decays to Z Boson Pairs,''
{\it Phys. Rev. Letters} {\bf 110} (2013) 081803.
\bibitem{ref:Bluj2013} Bluj M for the ATLAS and CMS Collaborations,
  ``Higgs Boson in the Standard Model and Other Highlights of SM Measurements with the LHC,''
  Acta Phys.\ Polon.\ B {\bf 44} (2013) 11,  2087.
\bibitem{ref:whm126}International Workshop, {\em Why $M_H=126$ GeV?}, Madrid,
  Spain, 25-27 Sep. 2013, {\tt http://workshops.ift.uam-csic.es/WMH126/index.html}
\bibitem{ref:alekhin12}Alekhin S, Djouadi A, Moch S, 
``The top quark and Higgs boson masses and the stability of the electroweak vacuum,''
{\it Physics Letters B} {\bf 716} (2012) 214-219.
\end{thebibliography}
\end{document}